\documentclass[twocolumn]{aastex63}
\usepackage{xcolor}
\usepackage{hyperref}

\received{Sep 09 2020 }
\revised{Dec 10 2020 }
\accepted{Jan 9, 2021}

\submitjournal{AJ}

\shorttitle{3FGL X-ray Analysis \& ML }
\shortauthors{Kerby et al.}

\graphicspath{{./}{/}}

\begin{document}

\title{X-ray Spectra and Multiwavelength Machine Learning Classification for Likely Counterparts to \textit{Fermi} 3FGL Unassociated Sources}

\author[0000-0003-2633-2196]{Stephen Kerby}
\affiliation{Department of Astronomy and Astrophysics \\
 Pennsylvania State University
University Park, PA 16802, USA}

\author[0000-0002-0878-1193]{Amanpreet Kaur}
\affiliation{Department of Astronomy and Astrophysics \\
 Pennsylvania State University
University Park, PA 16802, USA}

\author[0000-0002-5068-7344]{Abraham D. Falcone}
\affiliation{Department of Astronomy and Astrophysics \\
 Pennsylvania State University
University Park, PA 16802, USA}

\author[0000-0002-3019-4577]{Michael C. Stroh}
\affil{Center for Interdisciplinary Exploration and Research in Astrophysics (CIERA), Northwestern University, Evanston, IL 60201, USA}

\author{Elizabeth C. Ferrara}
\affil{NASA Goddard Space Flight Center, Greenbelt, MD 20771, USA}
\affil{Department of Astronomy, University of Maryland College Park, MD 20742, USA}

\author{Jamie A. Kennea}
\affiliation{Department of Astronomy and Astrophysics \\
 Pennsylvania State University
University Park, PA 16802, USA}

\author{Joseph Colosimo}
\affiliation{Department of Astronomy and Astrophysics \\
 Pennsylvania State University
University Park, PA 16802, USA}

\begin{abstract}
    We conduct X-ray spectral fits on 184 likely counterparts to \textit{Fermi}-LAT 3FGL unassociated sources.  Characterization and classification of these sources allows for more complete population studies of the high-energy sky. Most of these X-ray spectra are well fit by an absorbed power law model, as expected for a population dominated by blazars and pulsars. A small subset of 7 X-ray sources have spectra unlike the power law expected from a blazar or pulsar and may be linked to coincident stars or background emission.  We develop a multiwavelength machine learning classifier to categorize unassociated sources into pulsars and blazars using gamma- and X-ray observations.  Training a random forest procedure with known pulsars and blazars, we achieve a cross-validated classification accuracy of $98.6 \%$.  Applying the random forest routine to the unassociated sources returned 126 likely blazar candidates (defined as $P_{bzr} \ge 90 \% $) and 5 likely pulsar candidates ($P_{bzr} \le 10 \% $). Our new X-ray spectral analysis does not drastically alter the random forest classifications of these sources compared to previous works, but it builds a more robust classification scheme and highlights the importance of X-ray spectral fitting. Our procedure can be further expanded with UV, visual, or radio spectral parameters or by measuring flux variability.
\end{abstract}


\section{Introduction}
\label{sec:Intro}

The bulk of the $3033$ sources in the \textit{Fermi} Gamma-ray Space Telescope - Large Area Telescope (\textit{Fermi}-LAT) 3FGL catalog fall into one of two classes: extragalactic blazars or nearby pulsars  \citep{Acero2015}. A small number of other identified sources include supernova remnants, X-ray binaries, and starburst galaxies \citep{Ferrara2015}.  However, $1010$ 3FGL sources are ``unassociated", without a confident astrophysical identification with a known pinpointed source, or with several competing astronomical explanations within the gamma-ray detection area. Based on the dominance of blazars and pulsars among identified 3FGL sources, many of these unassociated sources are likely blazars and pulsars.  Capturing these unassociated and heretofore unclassified sources would create a more complete all-sky sample of various classes of objects, particularly blazars.

Categorizing the blazars and pulsars in the 3FGL unassociated list is an important step towards confident population studies of both classes.  The \textit{Fermi}-LAT unassociated sources might include blazars that are lower luminosity or higher redshift than their more easily detected and identified cousins in the established 3FGL blazar catalog \citep{Ferrara2015}.  Therefore, pursuing the classification of the 3FGL unassociated sources will help build a more complete population study, which will aid in verification and analysis of the blazar sequence \citep[e.g.,][]{Fossati1998,Ghisellini2017} as a theoretical unifying scheme for blazars.

In a similar way, identification and classification of 3FGL unassociated sources can also lead to new pulsar candidates, a population which has included new accreting pulsars in the 3FGL catalog \citep[e.g.][]{Wu2018,KwanLok2018}. Furthermore, in-depth analysis might show that an object suits neither blazar nor pulsar classification. Such sources might represent new gamma-ray binaries, or possibly more exotic objects \citep{SazParkinson2016}. The 3FGL unassociated list could contain several such objects, but classification of the  blazars and pulsars that probably make up most of the unassociated sample is the first step in identifying any unique sources.

While gamma-ray observations from \textit{Fermi}-LAT are the foundation for the 3FGL catalog, matching gamma-ray sources with X-ray counterparts and extending analysis to lower energy photons is a vital step in deeper analysis of the 3FGL sources. Recent work \citep{Kaur2019} developed a machine learning (ML) approach to sort 217 high-S/N unassociated 3FGL sources into blazars and pulsars. The 217 unassociated sources were analyzed by combining the gamma-ray flux and photon index of each object with a coarse estimate for the X-ray flux using probable counterpart X-ray excesses from the Neil Gehrels \textit{Swift} Observatory (aka \textit{Swift}) \citep{Gehrels2004}. For simplicity, the X-ray fluxes used in that work assumed an X-ray photon index of $\Gamma_{X} = 2$ rather than conducting full spectral fits.  Training an ML routine with known pulsar and blazar samples, the authors identified $173$ likely blazars with $P_{bzr} >90 \%$ ($134$ with $P_{bzr} >99 \%$) and $13$ likely pulsars with $P_{bzr} <10 \%$ ($7$ with $P_{bzr} < 1 \%$). $31$ sources from the 3FGL unassociated list defied categorization and were labeled 'ambiguous'.

The unassociated sources examined in \citep{Kaur2019} each had only one high-S/N X-ray excess in their gamma-ray confidence ellipse, and the majority of gamma-ray sources are expected to have counterpart X-ray sources. This is especially true of blazars, which make up the majority of the catalog.  In \cite{Kaur2019} and this work we only examine gamma-ray sources with a single high-S/N (S/N $>$ 4) X-ray excess in the confidence region, and for the purposes of classification, we make the initial assumption that this source is the counterpart. There may be some rare cases in which this X-ray source and the unassociated gamma-ray source do not correspond to one another, but since there is no other strong X-ray source in the Fermi error ellipse, this X-ray source is the most likely counterpart. 

In this work, we expand upon ML investigations by conducting detailed X-ray spectral analysis of 184 possible X-ray counterparts of the unassociated 3FGL sources. We obtain fully fitted X-ray fluxes and photon flux power law indices using an absorbed power-law model. Our training and validation sample was drawn from known lists of \textit{Fermi}-LAT blazars and pulsars which had data for all six gamma- and X-ray parameters used in our ML process \citep{Ackermann2015,Abdo2013}. \citet{Kaur2019} used a list of 217 unassociated sources for the test sample. 56 of those 217 objects later were associated with an astronomical object, so those were removed from consideration in this paper. Since then new observations have added 26 solitary X-ray excesses with high $S/N > 4$, leading to our initial list of 187 unassociated sources with one possible X-ray counterpart within the $95\%$ confidence region of the \textit{Fermi}-LAT unassociated source. We found three members of the 187 that were spurious contaminants from optically bright coincident stars creating optical loading in the \textit{Swift}-XRT detector, and we excluded these three excesses from all analysis.

The goal of this work is to conduct X-ray spectral analysis for some possible counterparts to \textit{Fermi}-LAT unassociated gamma-ray sources and to begin to build a multiwavelength classification routine. To this end we update the machine learning approach in \cite{Kaur2019} to include full X-ray spectral fits for high-S/N X-ray excesses spatially coincident with the 3FGL unassociated sources. In section \ref{sec:ObsAn}, we discuss observations, spectral fitting, and ML processes used in our analysis, plus we describe the training, test, and research samples.  Next, in section \ref{sec:Results} we tabulate and plot parameters for the various samples and we describe our fitting and classification results.  In section \ref{sec:DisCon} we discuss the classifications in comparison to previous works and summarize our findings.  Tables of spectral fits and ML classification results are also included.

\section{Observations and Analysis}
\label{sec:ObsAn}

\subsection{\textit{Swift}-XRT Observations of 3FGL Unassociated Sources}

Our sample is based on a collection that initially contained 187 \textit{Swift} X-ray counterparts to \textit{Fermi}-LAT 3FGL sources with high detection Signal-to-Noise (S/N) ratio ($S/N \ge4$) and only a single X-ray excess in the $95\%$ confidence region of the 3FGL source. As three of these counterparts (3FGL J0858.0$-$4843, J1050.6$-$6112, and J1801.5$-$7825) were near bright ($m_V<8$) stars, the X-rays from those excesses are almost certainly spurious products of optical loading\footnote{\url{https://www.swift.ac.uk/analysis/xrt/optical_loading.php}} in the \textit{Swift}-XRT (X-ray Telescope) \citep{Burrows2005} detector. We excluded these false detections from all further analysis. Particularly bright stars in the field of view cause optical loading in the \textit{Swift}-XRT as optical photons contaminate the detector and introduce spurious signals. Optical loading is less likely for dimmer coincident stars. 

We used the NASA HEASARC keyword interface to download all \textit{Swift}-XRT observations within 8' of the centroid position of each 3FGL counterpart.  Two more sources (3FGL J1216.6-0557 and 3FGL J0535.7-0617c) were matched with \textit{Swift}-XRT observations only upon expanding the HEASARC search radius to 10'. These sources are positioned close to the edge of the \textit{Swift}-XRT field of view in their respective observations.

In total we collected over $500$ individual \textit{Swift}-XRT observations, capturing all $184$ members of the unassociated list. All observations used the photon counting (PC) mode of the \textit{Swift}-XRT, enabling two-dimensional imaging across the 23 arcminute XRT field-of-view. Usable \textit{Swift}-XRT exposure times for most objects was around 4 ks, but ranged from 1 ks to over 60 ks.

We cleaned and processed each level 1 event file using \verb|xrtpipeline| v.0.13.5 from the HEASOFT software\footnote{\url{https://heasarc.gsfc.nasa.gov/docs/software.html}}, then merged with other observations of that particular object using \verb|xselect| v.2.4g and \verb|ximage| v.4.5.1 to create a single summed event list for each source plus a summed exposure map and ancillary response file using \verb|xrtmkarf|.  For each unassociated 3FGL object, we produced spectra for source and background regions using \verb|xselect|. The source region was circular with radius 20 arcseconds, and the background region was annular with inner and outer radii of 50 and 150 arcseconds. Both regions were centered on the coordinates of the examined X-ray excess. 

If the count rate in the source region of any excess exceeded 0.5 counts per second, we would draw a new annular source region with an inner radius depending on the count rate to avoid photon pile-up and saturation on the detector. The possible X-ray counterparts to 3FGL unassociated sources are faint enough that none caused pile-ups to warrant annular source regions. Adding X-ray variability to our spectral and photometric analysis of the X-ray excesses would certainly be an interesting probe into variability timescales of blazars and pulsars, but unfortunately most of the sources examined here do not have X-ray observations spanning a wide enough time range to maintain a large training, test, or research sample.

Finally, we examined the total number of counts in the source region to determine whether to use $\chi^2$ initial spectral fitting. The Cash statistic is a useful fitting statistic for spectra with few counts, particularly in cases for which there are not sufficient counts to group for a $\chi^2$ fit \citep{Cash1976}.  When an excess has enough photons in its source region, we binned our data with 20 counts per bin to enable an initial $\chi^2$ fit and also prepared a spectrum file for eventual Cash fitting. For an excess with only a few dozen detected X-ray photons, this approach would result in only one or two bins, and $\chi^2$ fitting would produce unreliable non-Gaussian distributed fits. The Cash statistic does not require any such binning, and therefore is often used in fitting faint X-ray spectra with few counts. While the Cash statistic cannot be directly considered as a measure of goodness of fit like $\chi^2$, a Cash statistic similar to the degrees of freedom is a rough indicator of a reasonable fit.
 
\subsection{Detailed X-ray Spectral Fitting}

While the machine learning analysis of the X-ray counterparts to 3FGL unassociated sources in \cite{Kaur2019} assumed an X-ray photon index $\Gamma_{X} = 2$ to calculate X-ray flux, our archival \textit{Swift}-XRT observations facilitated a full spectral fitting. We used \verb|Xspec| v.12.10.1f \citep{Arnaud1996} to fit each spectrum. The fitting model $tbabs \times cflux \times powerlaw$ included three nested functions: \verb|tbabs|, \verb|cflux|, and \verb|powerlaw|. \verb|cflux| calculated the total unabsorbed flux between 0.3 and 10 keV and \verb|tbabs| modeled line-of-sight hydrogen extinction using galactic values from the \verb|nH| lookup function described in \cite{Wilms2000}. 

The galactic line-of-sight extinction is fixed at the catalog value for each spectrum analyzed. \verb|powerlaw| is a simple power law. Uncertainties on the fitted photon index and X-ray flux were jointly measured using the iterative \verb|steppar| routine; this routine occasionally encountered numerical errors finding the error of photon indices close to zero. For these spectra we report the symmetric error generated by \verb|fit|. Spectra with high photon counts were initially fit using $\chi^2$ as the optimization statistic to create first guesses for Cash statistic fitting.

\begin{figure*}
    \centering
    \includegraphics[width=0.8\textwidth]{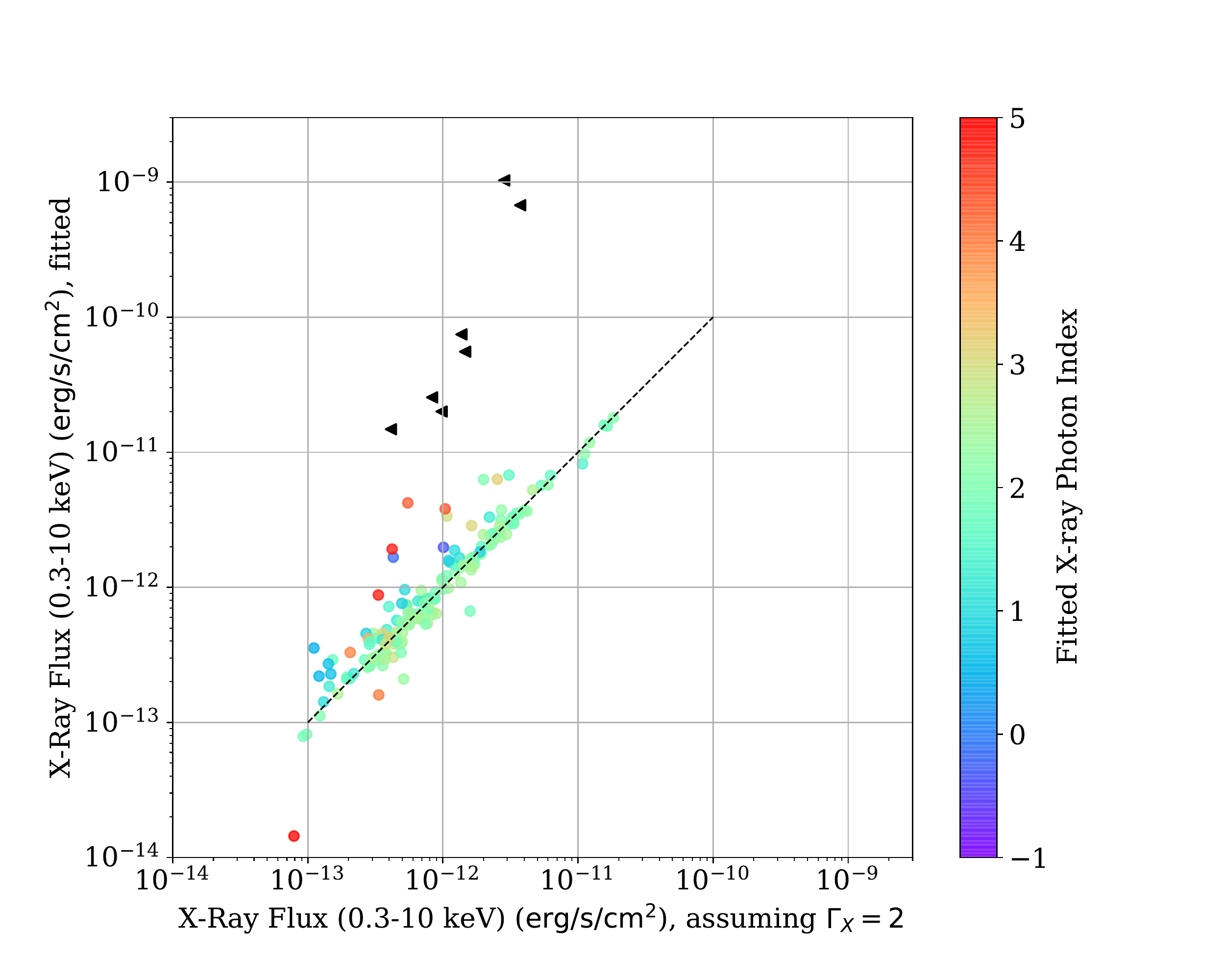}
    \caption{Fully fitted X-ray flux vs X-ray flux assuming $\Gamma_{X} = 2$. Points are colored based on fitted X-ray photon index. Black triangles are excesses identified as having extreme X-ray photon indices in table \ref{tab:trouble}. This figure excludes the five excesses for which full X-ray fitting failed to converge.}
    \label{fig:indexcheck}
\end{figure*}

Comparing the new fitted fluxes to fluxes calculated assuming $\Gamma_{X} = 2$ provides a useful sanity check for the X-ray spectral fitting routine.  Objects with fitted $\Gamma_X$ close to $2$ should show fitted X-ray fluxes close to the previous X-ray fluxes that assumed $\Gamma_{X} = 2$. Large revisions in X-ray flux should be reserved for objects with $\Gamma_{X} \ne 2$.  Figure \ref{fig:indexcheck} shows this reassuring close correspondence for spectra with fitted photon indices near $\Gamma_{X} = 2$, while spectra with fitted indices departing from $2$ show significantly corrected fluxes. All of the successful spectral fits are displayed in Table \ref{tab:fitresults}.

Of the eleven X-ray excesses whose fully fitted X-ray fluxes differed by more than an order of magnitude from the X-ray flux assuming $\Gamma_X = 2$, seven corresponded to the individually analyze unusual sources described in table \ref{tab:trouble}. Of the remaining four, one with a fitted index of $\Gamma_X = 4.4$ saw a drop in X-ray flux by a factor of ten compared to $\Gamma_X = 2$ fits. This gamma-ray source (3FGL J1837.3-2403) has \textit{Swift} exposures of such duration that at least four X-ray excesses are expected to spuriously appear in the gamma-ray confidence region, suggesting that this X-ray source is background. The remaining three X-ray excesses with radically different fitted fluxes are three of the five excesses where spectral fitting failed to converge, described above. Figure \ref{fig:indexcheck} excludes only the five excesses with unconverged X-ray fitting.

After fitting with a power-law model, seven excesses showed unusually high or low photon indices compared to the photon index of $\Gamma_X \sim 2$ expected for pulsars and blazars. We conducted further analysis on these excesses listed in Table \ref{tab:trouble}, and they were excluded from the main ML process since they warranted individual investigation to check for coincident stars or catalog objects that might explain their spectra.  Six of these seven excesses are located within a few arcseconds of dim catalogued stars in the galactic plane, possible targets for future observations to verify whether the X-ray radiation is linked to the star. A quantitative estimate of optical loading for \textit{Swift}-XRT showed that none of the seven spectra can have count rates heavily impacted by optical loading from the nearby star, although one of the seven could have a minor contribution.

\begin{deluxetable*}{crl}
\tablecaption{Seven of the possible X-ray counterparts to unassociated 3FGL sources with extreme X-ray photon indices, $\Gamma_X < -1$ or $\Gamma_X > 5$, were excluded from the main ML classification effort.  We describe the spectrum of the excess and list any notable objects close to the X-ray excess via SIMBAD and NASA Extragalactic Database coordinate searches. Possible stellar counterparts include spectral type and apparent magnitude from SIMBAD if available.}
\label{tab:trouble}
\tablewidth{0pt}
\tablehead{
\colhead{3FGL Source} & \colhead{$\Gamma_{X}$} & \colhead{Notes}}
\startdata
3FGL J0748.8$-$2208 & 6.29 & very few counts, 2" from TYC 5993-3722-1 ($m_V = 12.4$) \\
%3FGL J0858.0$-$4834 & 7.95 & few counts, flat spectrum, 3" from HD77020 (G8II, $m_V = 5.9$) \\
3FGL J0905.6$-$4917 & 7.84 & diffuse in XRT image, listed as `confused' in 4FGL, 3" from   2MASS J09053033-4918382  (M4, $m_J = 9.5$) \\
%3FGL J1050.6$-$6112 & 10.0 &  peaked spectrum at 0.8 keV, 4" from HD94144 (B4III, $m_V = 6.8$) \\
3FGL J1329.8$-$6109 & 6.22 &  peaked spectrum, 10" from HD117110 (G0V, $m_V = 9.2$) \\
3FGL J1624.1$-$4700 & 7.42 & peaked spectrum,  1" from CD-46 10711  (K1IV rotationally variable star, $m_V = 11.0$)\\
3FGL J1710.6$-$4317 & 6.32 &  peaked spectrum at 0.9 keV \\
%3FGL J1801.5$-$7825 & $-$1.11 & 1" from HD162298 (K4III, $m_V = 7.2$) \\
3FGL J1921.6+1934 & 6.19 & flat spectrum, 2" from HD231222 ($m_V = 10.8$) \\
3FGL J2035.8+4902 & 5.82 & peaked spectrum at 0.8 keV, 5" from V* V2552 Cyg (Eclipsing binary, $m_V = 10.8$) \\
\enddata
\end{deluxetable*}

\subsection{Machine Learning}

While multi-wavelength spectral analysis enables comprehensive study of individual unassociated gamma-ray sources, the observations and interpretation of hundreds of such objects would pose an onerous time burden on human scientists.  Fortunately, recent developments in machine learning (ML) techniques have resulted in numerous applications of ML classification schemes to \textit{Fermi}-LAT unassociated source catalogs \citep{Hassan2012, SazParkinson2016, McFadden2017}. These developments are part of a wave of ML techniques promulgating into survey analysis.

In this work, we use a random forest (RF) classifier, an aggregate of many individual decision tree (DT) realizations, to classify sources into blazars and pulsars, following a procedure described in \cite{Breiman2001}. Our approach here is nearly identical to that in \cite{Kaur2019}, which achieved $\approx 95 \%$ accuracy with a decision tree method and $\sim 99 \%$ accuracy with a random forest approach.

A detailed description of the statistics and theory of decision trees and random forests can be found in \cite{Breiman2001}. In brief, decision tree classifiers are non-parametric supervised and trained machine learning methods.  DT classifiers discriminate objects between classes by branching classes one by one at decision nodes, each node judging a single parameter of an object via an inequality. A tree is optimized using the Gini impurity index, representing the probability of a randomly selected source from the dataset being incorrectly labeled at one decision node.

The random forest approach compounds the DT method described above, generating a forest of decision trees and classifying test objects based on the average of multiple decision trees \citep{Breiman2001}. An RF algorithm constructs numerous decision trees by randomly creating subsamples of the training dataset. The overall forest also returns the relative importance of the parameters of the training dataset.

Once the forest is fully trained, a new observation is assigned a classification probability based on the average of the classifications of each tree in the forest.  Overall, the use of many decision trees in the RF routine creates a more robust analysis of test objects and prevents overfitting in a single tree from biasing results. The DT and RF methods used in this paper utilize the \verb|sklearn| package available in Python.

\subsection{Training and Test Samples}

To train the RF classifier, we gathered a sample of 831 known sources, including 772 known blazars and 59 known pulsars. The sample was derived from the 3FGL catalog which provided gamma-ray properties and second Swift X-ray Point Source Catalog \citep[2SXPS, ][]{Evans2020} which provided X-ray parameters. In addition, the X-ray properties of known pulsars were obtained from the literature search of various studies; \citep[e.g.,][]{Marelli2012,SazParkinson2016, Wu2018, Zyuzin2018}. The parameters for each source included\footnote{In this work, $\log{x}$ always refers to the logarithm in base 10 of $x$.}:
\begin{itemize}
    \item X-ray photon index $\Gamma_X$
    \item Gamma-ray photon index $\Gamma_\gamma$
    \item The logarithm of gamma-ray flux $\log{F_\gamma}$
    \item The logarithm of X-ray to gamma-ray flux ratio $\log{F_X/F_\gamma}$
    \item The significance of the curvature in the gamma-ray spectrum (henceforth simply \textit{curvature})
    \item The gamma-ray variability index
\end{itemize}
All the X-ray parameters were determined through our analysis as explained earlier, while the gamma ray parameters were extracted from the 3FGL catalog. The complete details of the obtained gamma-ray data and the methods are provided in \citet{Kaur2019}.

Because there are many more known blazars than known pulsars in the training and test samples, we used Synthetic Minority Over-sampling Technique (SMOTE) \citep{Chawla2002} to generate synthetic members of the underrepresented class (pulsars) with a k-nearest neighbors approach.  Previous classification efforts have shown that seriously unbalanced training datasets can lead to trained RF classifiers that are biased against the underrepresented class \citep{Last2017}. The result of the SMOTE expansion is a catalog of known blazars and pulsars, plus artificial pulsars generated with the same distribution in parameter-space as the real pulsars, producing a catalog with a balanced number of 772 blazars and 772 pulsars. Expansion of the pulsar catalog via SMOTE is executed before the catalog is split into training and validation samples.

To optimize RF parameters such as number of individual trees and maximum tree depth, we utilized \texttt{GridSearchCV} in \texttt{sklearn v.0.20.3}. We found that 1000 decision trees splitting to a maximum depth of 15 nodes with at least one source in each leaf were required to effectively train the classifier. The reported blazar probability for each source is the fraction of the 1000 trees in which the object was classified as a blazar.

In this paper, we utilized a cross-validation method, \texttt{cross\_val\_predict} from \texttt{sklearn v.0.20.3} by dividing our total sample into 10 folds and then used each fold (one at a time) as a test sample For the validation step, a RF classifier is generated using a training subsample. Members of the corresponding test subsample are then classified as a pulsar or a blazar and the generated classification is compared to the actual class label. In this way, an accuracy score for the entire RF tree is generated. This process is repeated ten time, so ten random forest classifiers are trained and each is validated in turn; the overall accuracy of the RF classifier is the average of the validation accuracy of the ten folded iterations. The overall RF accuracy obtained in this way was 98.5\%. 

In \citet{Kaur2019}, the authors separated 100 sources from the complete set of blazars and pulsars for a test sample to calculate the accuracy of the classifier trained on the rest of the data set. This unitary test sample leads to an accuracy based on only that one test sample. In this way, the reported RF validation accuracy in that paper measures the same reliability as in this paper, but with a more restricted approach to selecting a test sample.

\begin{deluxetable*}{cccccc}
\tablecaption{RF feature importance}
\label{tab:importan}
\tablewidth{0pt}
\tablehead{
\colhead{$\Gamma_X$} & \colhead{$\Gamma_\gamma$} & \colhead{Curvature} & \colhead{Variability index} & \colhead{$\log{F_\gamma}$} & \colhead{$\log{F_X/F_\gamma}$}
}
\startdata
0.033 & 0.097 & 0.41 & 0.13 & 0.08 & 0.25 \\
\enddata
\end{deluxetable*}

\subsection{Unassociated Sample}

We conducted an initial investigation of our sample by combining gamma-ray properties from the 3FGL catalog for the 184 unassociated sources with the X-ray properties derived from the new spectral fits. Comparing the photometric and spectroscopic properties of the unassociated sample with those of the known pulsars and blazars, the unassociated sources tend to have lower gamma-ray fluxes than both the known blazars and pulsars. The mean X-ray fluxes of the unassociated counterparts fall between the mean fluxes of the known blazars and known pulsars.

\begin{figure*}
    \centering
    \includegraphics[width=1.0\textwidth]{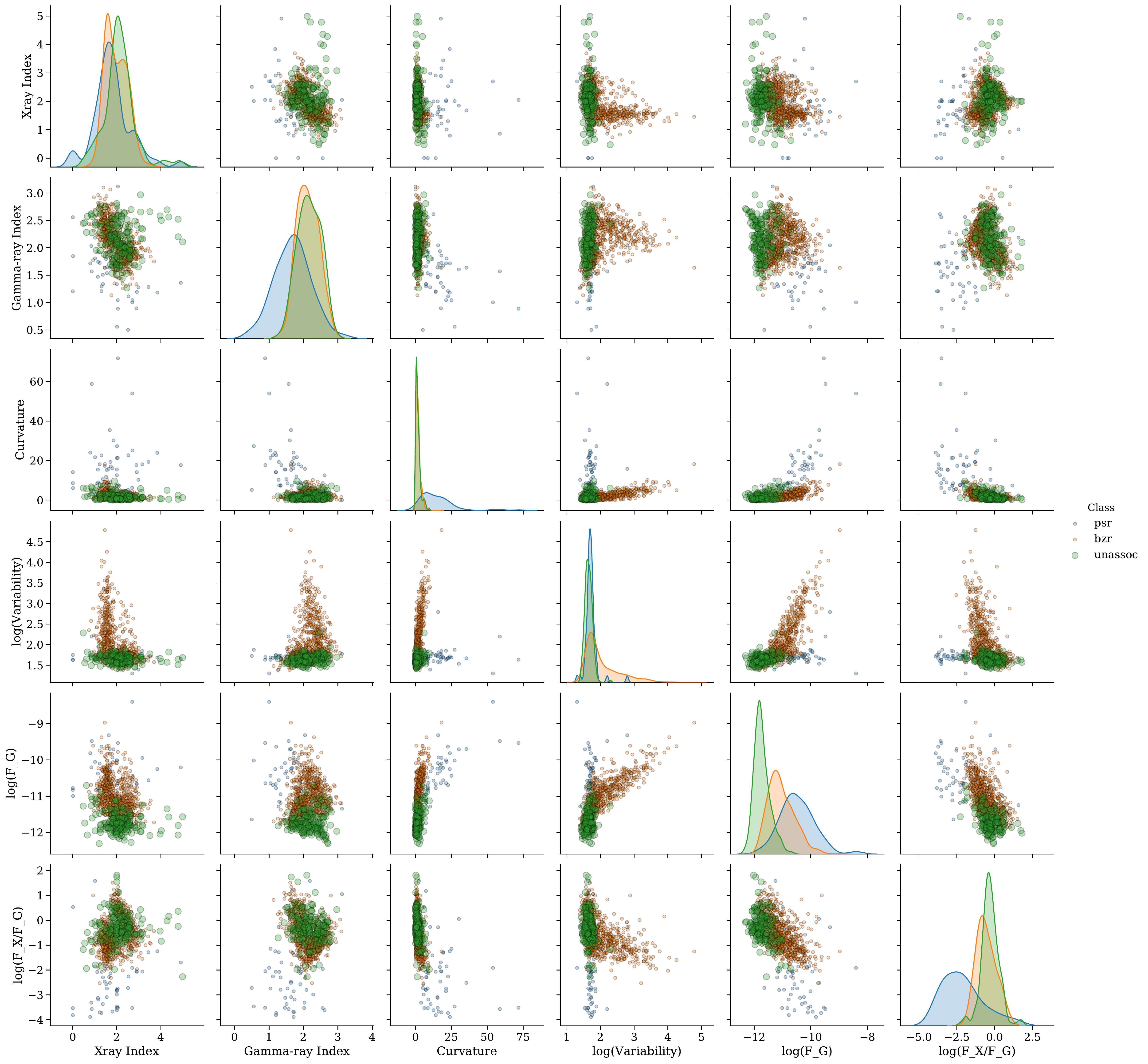}
    \caption{A full pairs plot of the known blazars (red), known pulsars (blue) and the unassociated sources (green). Histograms in the diagonal plots are smoothed and normalized to different scales for each class. The six parameters are the X-ray photon index, the gamma-ray photon index, the gamma-ray curvature index, the logarithm of the variability index, the logarithm of the gamma-ray flux in $\rm{erg/s/cm^2}$, and the logarithm of the ratio of X-ray to gamma-ray flux}
    \label{fig:Pairs}
\end{figure*}

The histograms in Figure \ref{fig:Pairs} show that the unassociated sources most readily overlap with the known blazar sample, suggesting that the majority of the unassociated sources should be blazars, consistent with the membership of known 3FGL sources.  Interestingly, the histogram for X-ray photon index (the plot in the upper-left corner of Figure \ref{fig:Pairs}) shows two distinct peaks in the known 3FGL blazar distribution, with the unassociated source distribution overlapping primarily with the higher/softer peak.

Our model incorporating hydrogen-extincted power law spectra returned unusually high or low $\Gamma_X$ for seven sources, as described above and listed in Table \ref{tab:trouble}.  Given that some of these excesses also coincided with catalogued stars, we view these sources as dubious and do not include these seven sources in the ML classification.

\section{Results}
\label{sec:Results}

Our X-ray spectral fits and RF classification results for the entire 3FGL unassociated catalog is available at CDS via anonymous ftp to cdsarc.u-strasbg.fr (130.79.128.5) or via \href{http://cdsarc.u-strasbg.fr/viz-bin/qcat?J/AJ}{http://cdsarc.u-strasbg.fr/viz-bin/qcat?J/AJ}

The vast majority of examined X-ray excesses obtained well-defined spectral fits, reported in Table \ref{tab:fitresults}. Most of the fits to the X-ray spectra have photon indices between $0$ and $4$, a similar range to the lists of known pulsars and blazars used to train the RF routine (as in Figure \ref{fig:Pairs}). This supports our first assumption that most of the unassociated sources are pulsars or blazars.  These fits represent a large collection of X-ray parameters for likely counterparts to previously unassociated 3FGL gamma-ray sources. Five excesses had very few X-ray photons after summing up the \textit{Swift}-XRT observations.  With so few photons, the \verb|xspec| fitting routine could not return useful spectral fits. For these spectra, we assumed $\Gamma_{X} = 2$ to calculate the flux. 

The importances of the different parameters in the RF classifier indicate the features of the gamma- and X-ray spectra that are the strongest predictors of blazar or pulsar identification. The two most important features are gamma-ray spectral curvature and $\log{F_X/F_\gamma}$, shown in Table \ref{tab:importan}. Figure \ref{fig:indexcheck} shows that full X-ray fitting with both flux and photon index as free parameters did not alter X-ray flux by more than an order of magnitude for the vast majority of spectra. Some excesses did see a change in X-ray flux by around a half an order of magnitude; these spectra also showed the largest alterations in fitted photon index from $\Gamma_X \sim 2$. By this measure, fully fitting X-ray spectra instead of assuming $\Gamma_X = 2$ can correct reported X-ray flux by up to an order of magnitude and obtain photon index as a fitted parameter.

Applying the optimized RF classifier to the 177 fitted unassociated sources and X-ray excesses (ignoring the seven troublesome spectra discussed above), we use cross-validated blazar probabilities to categorize the unassociated sources. In this way, we use each of the ten subfolds used to validate the RF accuracy and average the blazar probability of each source from each fold. We identified 5 likely pulsars ($P_{bar} \le 10 \%$) and 126 likely blazars ($P_{bar} \ge 90 \%$), with 46 sources remaining ambiguous. The results from this classification are reported in Table \ref{tab:MLresults}.

\begin{figure*}
    \centering
    \includegraphics[width=0.6\textwidth]{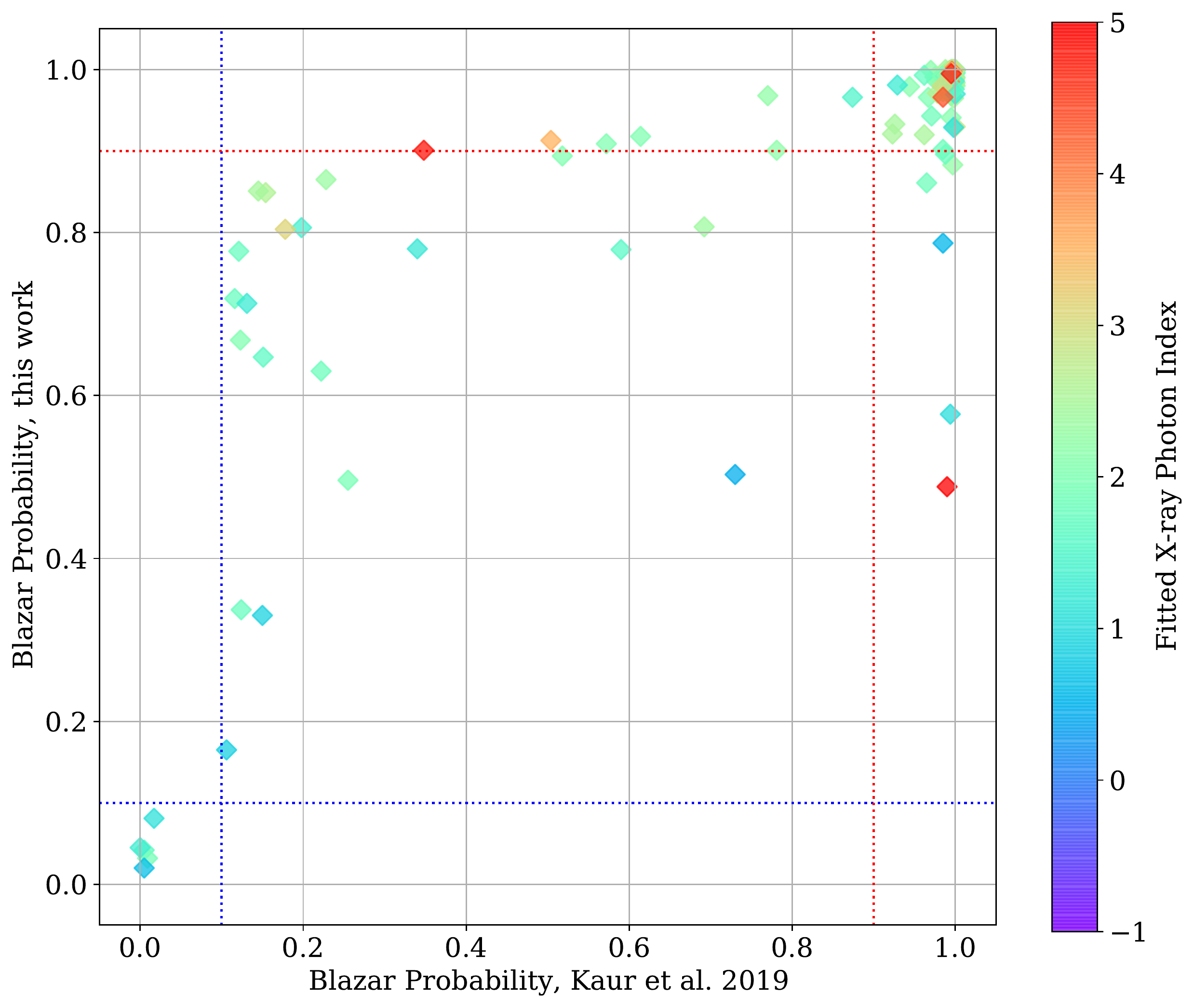}
    \caption{Blazar probability in this work vs in \cite{Kaur2019} for excesses analyzed in both works.  Dotted red and blue vertical and horizontal lines show $<10\%$ (likely pulsar) and $>90\%$ (likely blazar) categorization bounds respectively. Points are color-coded by fully fitted X-ray photon indices with the same scale as figure \ref{fig:indexcheck}}
    \label{fig:OldNew}
\end{figure*}

Figure \ref{fig:OldNew} compares the blazar probability for the RF classification in this work to the same in \cite{Kaur2019} for the $161$ sources analyzed in both works, color-coding the points by the fully fitted X-ray photon index. While the validation accuracy of the new RF classifier is not significantly different from the approach in \cite{Kaur2019}, we have refined the blazar and pulsar catalogs and introduced new spectral information to all sources by fully fitting for photon index and X-ray flux. These alterations suggest that the new $P_{bzr}$ values are more reliable than previous versions, facilitating the direct comparison in Figure \ref{fig:OldNew}.

In general, the addition of fully-fitted X-ray indices and fluxes to the ML training and test data sets did not alter the RF classifications for most of the 3FGL unassociated sources, and there is no pattern linking severe alterations in flux or photon index to previous estimates with drastically changed blazar probabilities.  The most significant change in classifications were 13 sources previously classified as likely blazars and 3 previously classified as likely pulsars but here were labeled as ambiguous.  Additionally, 7 ambiguous classifications in that previous work were here labeled as likely blazars. 138 of the 161 shared sources were sorted into the same category in both approaches.

We did not see any systematic relation that could be attributed as arising from a one-to-one relationship between altered X-ray photon index and changes in $P_{bzr}$. While many sources were classified with a similar blazar probability in this work and in \citep{Kaur2019}, some were classified as blazars or pulsars with greater confidence with more comprehensive spectral fits. Figure \ref{fig:OldNew} shows that changes in blazar probability from \citep{Kaur2019} to this work occur independently of divergence from coarse estimated spectral parameters. However, there are three overarching trends diverging from a one-to-one correspondence in the blazar probabilities shown in figure \ref{fig:OldNew}. Some likely blazars became ambiguous (or vice versa), and there is a general shift among previously more ambiguous sources towards higher blazar probabilites.  That only ambiguous sources saw large systematic shifts towards higher blazar probabilities is reassurance that the addition of more comprehensive X-ray spectral fits adds valuable information for discriminating pulsars and blazars.

\begin{figure*}
    \centering
    \includegraphics[width=0.9\textwidth]{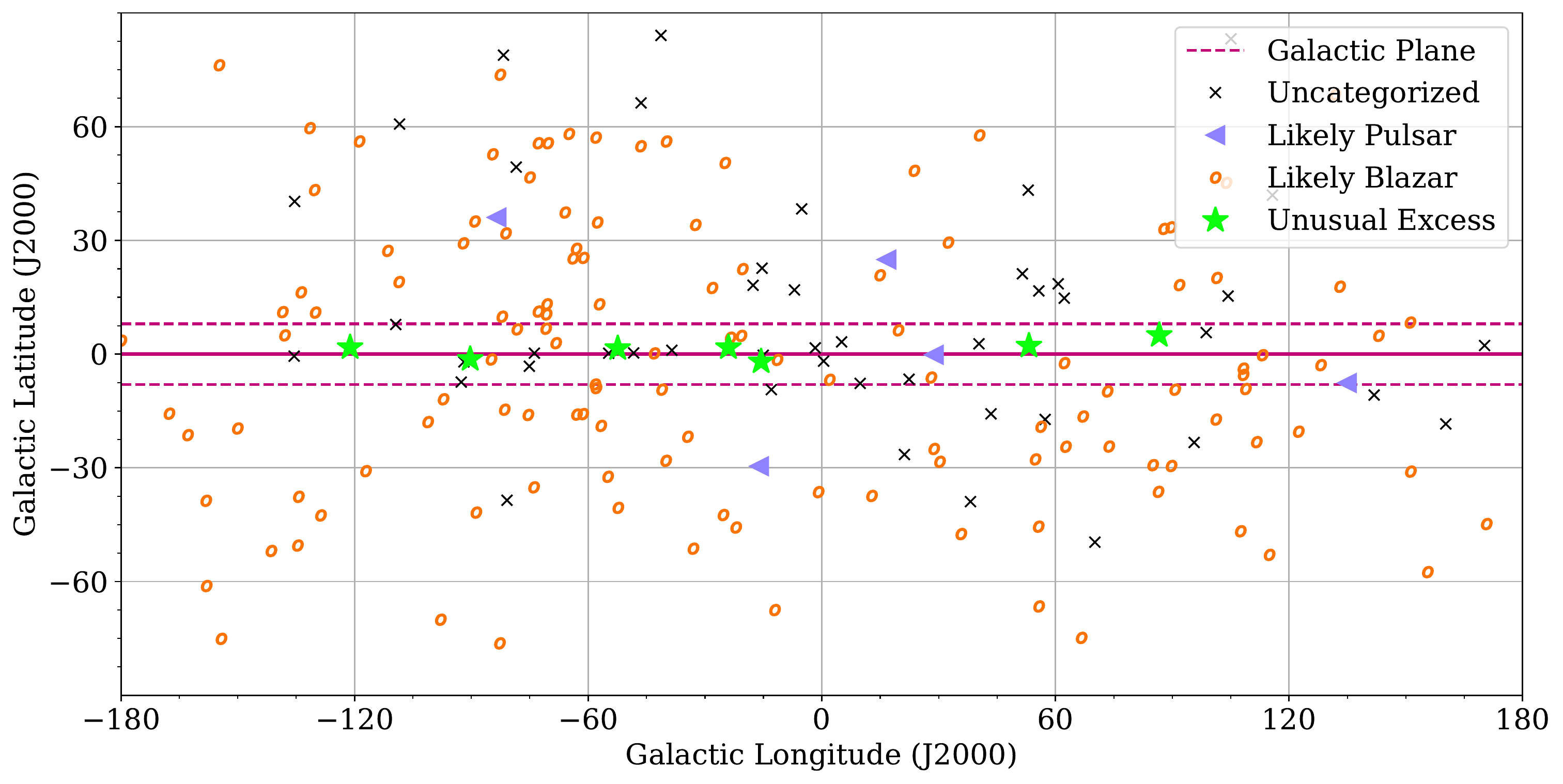}
    \caption{Galactic coordinates for the 184 X-ray counterparts examined in this work. Points include likely blazars, likely pulsars, and ambiguous counterparts, as well as the seven sources with unusual spectral fits.  The  galactic plane is approximately within the purple boundaries, and the galactic center is at $(l,b) = (0,0)$.}
    \label{fig:Coords}
\end{figure*}

The locations of the X-ray counterparts in galactic coordinates are shown in Figure \ref{fig:Coords}. While the counterparts classified as likely blazars and ambiguous are scattered roughly uniformly across the sky, the excesses with photon indices very different from expected for pulsars or blazars are almost entirely restricted to the galactic plane, which could be an indication that the X-rays of these excesses may be of galactic origin. If all the unusual excesses originate in similar astronomical objects, a catalog of gamma- and X-ray parameters of such objects could be used to add additional class to an RF scheme. As we do not know of a unified explanation for those excesses and as there is no previous catalogue of known similar objects it is not feasible to include sources of this unusual character as a class in the RF classifier method.

\section{Discussion and Conclusions}
\label{sec:DisCon}

In this work, we conducted full X-ray spectral analysis of counterparts to 3FGL unassociated sources, obtaining fluxes and photon indices for most of the X-ray sources examined. The vast majority of the X-ray sources linked to the 3FGL catalog unassociated sources were ably fit by our model, and represent a significant survey of dim excesses in the X-ray sky. Comprehensive X-ray spectral fits and the exclusion of unsuitable spectra together increase confidence in the output of the ML classification.

After training the RF classifier, the feature importances show in Table \ref{tab:importan} show that $log(F_X/F_G)$ was heavily weighted compared to most other parameters in the ML process, indicating that the ratio is an important discriminator for discerning blazars from pulsars. With previous estimates for X-ray flux differing from fully-fit fluxes by a factor of two or more, full X-ray spectral analysis is an important contribution to ML classification of pulsars and blazars and to cataloging unassociated source X-ray parameters.

Finally, we compared our classification results to those obtained in previous work and found that introducing fully fit X-ray parameters achieves the same RF classification accuracy ($98.5 \%$ in this work) as in \cite{Kaur2019} ($99\%$) where the X-ray flux was obtained by assuming $\Gamma_{X} = 2$. The new approach with full X-ray fitting essentially matches the validation accuracy of the fixed photon index approach in \cite{Kaur2019}, fulfilling our goal of meeting the old validation accuracy as a target. 138 of the 161 sources examine in both investigations were classified similarly. Including full X-ray spectral parameters shifts some sources into or out of the ``likely blazar" category and shifts previously ambiguous sources to higher blazar probability.

Figure \ref{fig:OldNew} shows a general increase in blazar probability for previously ambiguous sources, while likely pulsars did not see an increase in blazar probability. Given that many of the ambiguous sources are probably blazars, this trend suggests that the addition of X-ray spectral fits adds valuable information for discerning pulsars from blazars even if the changes are not dramatic enough to shift sources to 100\% blazar probability.

In conducting full X-ray fitting, we discovered several clearly spurious X-ray sources. The elimination of these excesses from consideration is an important step in ensuring a reliable classification method, as those excesses would otherwise remain in consideration for classification as blazars or pulsars. By identifying possible stellar or instrumental origins for seven sources with X-ray photon indices $\Gamma_X<-1$ or $\Gamma_X>5$, we showed that X-ray fitting can also sift out excesses that should not be immediately classified into blazars or pulsars via ML, further increasing the reliability of our results.

Besides a few sources identified as optical loading contaminants due to nearby bright stars, we selected seven spectra with X-ray photon indices significantly divergent from typical theoretical predictions of pulsars and blazars for further investigation.  Because these sources are largely contained within the galactic plane, it is likely that they originate within the Milky Way. That six of the seven have a star within a few arcseconds suggests that some of these excesses may be linked to stellar phenomena and may be interesting targets for future investigations. Alternatively, the nearby stars in these seven cases may have been simple coincidences that are unrelated to the X-ray or gamma-ray source. In one of the seven cases, the stars may be contributing some optical loading in the \textit{Swift}-XRT CCD, but for this $m_V=9.2$ star, the optical loading contribution is expected to be a small fraction relative to the detected X-ray count rate.

Though the requirements for X-ray fits create more stringent observational requirements in a multiwavelength ML routine compared to using only gamma-ray observations, incorporating diverse observations increases the comprehensive capabilities of the ML routine.  Assuming $\Gamma_{X} = 2$ to obtain X-ray flux admitted to the ML analysis spectra that now are revealed to be poorly described by a blazar or pulsar model. These spectra  might not actually be pulsars and blazars, but in assuming an X-ray photon index they might be incorrectly categorized. These unusual sources can be examined individually instead of analyzed automatically, allowing for more specialized future studies.

Even more comprehensive classification could be achieved by including X-ray variability measurements or expanding analysis into the ultraviolet bands using the UVOT telescope onboard \textit{Swift}  \citep{Roming2005}.  Analysis of lower-energy photons together with X-ray and gamma-ray photons could also constrain the location of the synchrotron peak in each spectrum, greatly increasing the detail in the characterization of the source. It will be particularly interesting to apply a growing understanding of pulsar and blazar classification to the results of surveys using new and upcoming space telescopes such as eROSITA. Whether the characteristics of bright pulsars and blazars are similar to those of dimmer pulsars and blazars should become evident as wide-field surveys extend to dimmer magnitudes.

Eventually, after the easier to find blazars and pulsars have been identified, it may become prudent to expand ML classification beyond a binary choice of pulsars and blazars. To that end, switching from a RF routine to a more detailed approach such as guided clustering would dramatically increase the flexibility of the automated system at the cost of more strenuous supervision requirements.

\section*{Acknowledgements}

We are grateful to the reviewer for their insightful and detailed commentary, which was vital in improving and focusing this work.

This research has made use of data and/or software provided by the High Energy Astrophysics Science Archive Research Center (HEASARC), which is a service of the Astrophysics Science Division at NASA/GSFC. The authors gratefully acknowledge the support of NASA grants 80NSSC17K0752 and 80NSSC18K1730.

Michael C. Stroh is partially supported by the Heising-Simons Foundation under grant 2018-0911.

\section*{Catalog}

The machine-readable tables corresponding to Tables \ref{tab:fitresults} and \ref{tab:MLresults} are available at CDS via anonymous ftp to cdsarc.u-strasbg.fr (130.79.128.5) or via \href{http://cdsarc.u-strasbg.fr/viz-bin/qcat?J/AJ}{http://cdsarc.u-strasbg.fr/viz-bin/qcat?J/AJ}. 

Additional thanks is due to the CDS and VizieR teams for facilitating this web service. 

\bibliography{/main}{}

\begin{thebibliography}{}
\expandafter\ifx\csname natexlab\endcsname\relax\def\natexlab#1{#1}\fi
\providecommand{\url}[1]{\href{#1}{#1}}
\providecommand{\dodoi}[1]{doi:~\href{http://doi.org/#1}{\nolinkurl{#1}}}
\providecommand{\doeprint}[1]{\href{http://ascl.net/#1}{\nolinkurl{http://ascl.net/#1}}}
\providecommand{\doarXiv}[1]{\href{https://arxiv.org/abs/#1}{\nolinkurl{https://arxiv.org/abs/#1}}}

\bibitem[{{Abdo} {et~al.}(2013){Abdo}, {Ajello}, {Allafort}, {Baldini},
  {Ballet}, {Barbiellini}, {Baring}, {Bastieri}, {Belfiore}, {Bellazzini}, \&
  et~al.}]{Abdo2013}
{Abdo}, A.~A., {Ajello}, M., {Allafort}, A., {et~al.} 2013, \apjs, 208, 17,
  \dodoi{10.1088/0067-0049/208/2/17}

\bibitem[{Acero {et~al.}(2015)Acero, Ackermann, Ajello, Albert, Atwood,
  Axelsson, Baldini, Ballet, Barbiellini, Bastieri, Belfiore, Bellazzini,
  Bissaldi, Blandford, Bloom, Bogart, Bonino, Bottacini, Bregeon, Britto,
  Bruel, Buehler, Burnett, Buson, Caliandro, Cameron, Caputo, Caragiulo,
  Caraveo, Casandjian, Cavazzuti, Charles, Chaves, Chekhtman, Cheung, Chiang,
  Chiaro, Ciprini, Claus, Cohen-Tanugi, Cominsky, Conrad, Cutini, D'Ammando,
  Angelis, Deklotz, Palma, Desiante, Digel, Venere, Drell, Dubois, Dumora,
  Favuzzi, Fegan, Ferrara, Finke, Franckowiak, Fukazawa, Funk, Fusco, Gargano,
  Gasparrini, Giebels, Giglietto, Giommi, Giordano, Giroletti, Glanzman,
  Godfrey, Grenier, Grondin, Grove, Guillemot, Guiriec, Hadasch, Harding, Hays,
  Hewitt, Hill, Horan, Iafrate, Jogler, Jóhannesson, Johnson, Johnson,
  Johnson, Johnson, Kamae, Kataoka, Katsuta, Kuss, Mura, Landriu, Larsson,
  Latronico, Lemoine-Goumard, Li, Li, Longo, Loparco, Lott, Lovellette,
  Lubrano, Madejski, Massaro, Mayer, Mazziotta, McEnery, Michelson, Mirabal,
  Mizuno, Moiseev, Mongelli, Monzani, Morselli, Moskalenko, Murgia, Nuss, Ohno,
  Ohsugi, Omodei, Orienti, Orlando, Ormes, Paneque, Panetta, Perkins,
  Pesce-Rollins, Piron, Pivato, Porter, Racusin, Rando, Razzano, Razzaque,
  Reimer, Reimer, Reposeur, Rochester, Romani, Salvetti, Sánchez-Conde,
  Parkinson, Schulz, Siskind, Smith, Spada, Spandre, Spinelli, Stephens,
  Strong, Suson, Takahashi, Takahashi, Tanaka, Thayer, Thayer, Thompson,
  Tibaldo, Tibolla, Torres, Torresi, Tosti, Troja, Klaveren, Vianello, Winer,
  Wood, Wood, \& Zimmer}]{Acero2015}
Acero, F., Ackermann, M., Ajello, M., {et~al.} 2015, Astrophysical Journal,
  Supplement Series, 218, \dodoi{10.1088/0067-0049/218/2/23}

\bibitem[{Ackermann {et~al.}(2015)Ackermann, Ajello, Atwood, Baldini, Ballet,
  Barbiellini, Bastieri, Gonzalez, Bellazzini, Bissaldi, Blandford, Bloom,
  Bonino, Bottacini, Brandt, Bregeon, Britto, Bruel, Buehler, Buson, Caliandro,
  Cameron, Caragiulo, Caraveo, Carpenter, Casandjian, Cavazzuti, Cecchi,
  Charles, Chekhtman, Cheung, Chiang, Chiaro, Ciprini, Claus, Cohen-Tanugi,
  Cominsky, Conrad, Cutini, D'Abrusco, D'Ammando, Angelis, Desiante, Digel,
  Venere, Drell, Favuzzi, Fegan, Ferrara, Finke, Focke, Franckowiak, Fuhrmann,
  Fukazawa, Furniss, Fusco, Gargano, Gasparrini, Giglietto, Giommi, Giordano,
  Giroletti, Glanzman, Godfrey, Grenier, Grove, Guiriec, Hewitt, Hill, Horan,
  Itoh, Jóhannesson, Johnson, Johnson, Kataoka, Kawano, Krauss, Kuss, Mura,
  Larsson, Latronico, Leto, Li, Li, Longo, Loparco, Lott, Lovellette, Lubrano,
  Madejski, Mayer, Mazziotta, McEnery, Michelson, Mizuno, Moiseev, Monzani,
  Morselli, Moskalenko, Murgia, Nuss, Ohno, Ohsugi, Ojha, Omodei, Orienti,
  Orlando, Paggi, Paneque, Perkins, Pesce-Rollins, Piron, Pivato, Porter,
  Rainò, Rando, Razzano, Razzaque, Reimer, Reimer, Romani, Salvetti, Schaal,
  Schinzel, Schulz, Sgrò, Siskind, Sokolovsky, Spada, Spandre, Spinelli,
  Stawarz, Suson, Takahashi, Takahashi, Tanaka, Thayer, Thayer, Tibaldo,
  Torres, Torresi, Tosti, Troja, Uchiyama, Vianello, Winer, Wood, \&
  Zimmer}]{Ackermann2015}
Ackermann, M., Ajello, M., Atwood, W.~B., {et~al.} 2015, Astrophysical Journal,
  810, \dodoi{10.1088/0004-637X/810/1/14}

\bibitem[{Arnaud(1996)}]{Arnaud1996}
Arnaud, K.~A. 1996, ASP Conference Series, 101

\bibitem[{Breiman(2001)}]{Breiman2001}
Breiman, L. 2001, 45, 5

\bibitem[{{Burrows} {et~al.}(2005){Burrows}, {Hill}, {Nousek}, {Kennea},
  {Wells}, {Osborne}, {Abbey}, {Beardmore}, {Mukerjee}, {Short}, {Chincarini},
  {Campana}, {Citterio}, {Moretti}, {Pagani}, {Tagliaferri}, {Giommi},
  {Capalbi}, {Tamburelli}, {Angelini}, {Cusumano}, {Br{\"a}uninger}, {Burkert},
  \& {Hartner}}]{Burrows2005}
{Burrows}, D.~N., {Hill}, J.~E., {Nousek}, J.~A., {et~al.} 2005, \ssr, 120,
  165, \dodoi{10.1007/s11214-005-5097-2}

\bibitem[{Cash(1976)}]{Cash1976}
Cash, W. 1976, Astronomy \& Astrophysics, 52, 307

\bibitem[{Chawla {et~al.}(2002)Chawla, Bowyer, Hall, \&
  Kegelmeyer}]{Chawla2002}
Chawla, N.~V., Bowyer, K.~W., Hall, L.~O., \& Kegelmeyer, W.~P. 2002, Journal
  of Artificial Intelligence Research, 16, 321

\bibitem[{Evans {et~al.}(2020)Evans, Page, Osborne, Beardmore, Willingale,
  Burrows, Kennea, Perri, Capalbi, Tagliaferri, \& Cenko}]{Evans2020}
Evans, P.~A., Page, K.~L., Osborne, J.~P., {et~al.} 2020, The Astrophysical
  Journal Supplement Series, 247, 54, \dodoi{10.3847/1538-4365/ab7db9}

\bibitem[{{Ferrara} {et~al.}(2015){Ferrara}, {Mirabal}, \& {Fermi-LAT
  Collaboration}}]{Ferrara2015}
{Ferrara}, E.~C., {Mirabal}, N.~R., \& {Fermi-LAT Collaboration}. 2015, in
  American Astronomical Society Meeting Abstracts, Vol. 225, American
  Astronomical Society Meeting Abstracts \#225, 336.02

\bibitem[{{Fossati} {et~al.}(1998){Fossati}, {Maraschi}, {Celotti}, {Comastri},
  \& {Ghisellini}}]{Fossati1998}
{Fossati}, G., {Maraschi}, L., {Celotti}, A., {Comastri}, A., \& {Ghisellini},
  G. 1998, \mnras, 299, 433, \dodoi{10.1046/j.1365-8711.1998.01828.x}

\bibitem[{{Gehrels} {et~al.}(2004){Gehrels}, {Chincarini}, {Giommi}, {Mason},
  {Nousek}, {Wells}, {White}, {Barthelmy}, {Burrows}, {Cominsky}, {Hurley},
  {Marshall}, {M{\'e}sz{\'a}ros}, {Roming}, {Angelini}, {Barbier}, {Belloni},
  {Campana}, {Caraveo}, {Chester}, {Citterio}, {Cline}, {Cropper}, {Cummings},
  {Dean}, {Feigelson}, {Fenimore}, {Frail}, {Fruchter}, {Garmire}, {Gendreau},
  {Ghisellini}, {Greiner}, {Hill}, {Hunsberger}, {Krimm}, {Kulkarni}, {Kumar},
  {Lebrun}, {Lloyd-Ronning}, {Markwardt}, {Mattson}, {Mushotzky}, {Norris},
  {Osborne}, {Paczynski}, {Palmer}, {Park}, {Parsons}, {Paul}, {Rees},
  {Reynolds}, {Rhoads}, {Sasseen}, {Schaefer}, {Short}, {Smale}, {Smith},
  {Stella}, {Tagliaferri}, {Takahashi}, {Tashiro}, {Townsley}, {Tueller},
  {Turner}, {Vietri}, {Voges}, {Ward}, {Willingale}, {Zerbi}, \&
  {Zhang}}]{Gehrels2004}
{Gehrels}, N., {Chincarini}, G., {Giommi}, P., {et~al.} 2004, \apj, 611, 1005,
  \dodoi{10.1086/422091}

\bibitem[{Ghisellini {et~al.}(2017)Ghisellini, Righi, Costamante, \&
  Tavecchio}]{Ghisellini2017}
Ghisellini, G., Righi, C., Costamante, L., \& Tavecchio, F. 2017,
  \dodoi{10.1093/mnras/stx806}

\bibitem[{Hassan {et~al.}(2012)Hassan, Mirabal, Contreras, \& Oya}]{Hassan2012}
Hassan, T., Mirabal, N., Contreras, J.~L., \& Oya, I. 2012, Monthly Notices of
  the Royal Astronomical Society, 428, 220, \dodoi{10.1093/mnras/sts022}

\bibitem[{Kaur {et~al.}(2019)Kaur, Falcone, Stroh, Kennea, \&
  Ferrara}]{Kaur2019}
Kaur, A., Falcone, A.~D., Stroh, M.~D., Kennea, J.~A., \& Ferrara, E.~C. 2019,
  \dodoi{10.3847/1538-4357/ab4ceb}

\bibitem[{{Last} {et~al.}(2017){Last}, {Douzas}, \& {Bacao}}]{Last2017}
{Last}, F., {Douzas}, G., \& {Bacao}, F. 2017, arXiv e-prints,
  arXiv:1711.00837.
\newblock \doarXiv{1711.00837}

\bibitem[{{Li} {et~al.}(2018){Li}, {Hou}, {Strader}, {Takata}, {Kong},
  {Chomiuk}, {Swihart}, {Hui}, \& {Cheng}}]{KwanLok2018}
{Li}, K.-L., {Hou}, X., {Strader}, J., {et~al.} 2018, \apj, 863, 194,
  \dodoi{10.3847/1538-4357/aad243}

\bibitem[{{Marelli}(2012)}]{Marelli2012}
{Marelli}, M. 2012, arXiv e-prints, arXiv:1205.1748.
\newblock \doarXiv{1205.1748}

\bibitem[{McFadden {et~al.}(2017)McFadden, Karastergiou, \&
  Roberts}]{McFadden2017}
McFadden, R., Karastergiou, A., \& Roberts, S. 2017, Proceedings of the
  International Astronomical Union, 13, 372–373,
  \dodoi{10.1017/S1743921317009000}

\bibitem[{{Roming} {et~al.}(2005){Roming}, {Kennedy}, {Mason}, {Nousek}, {Ahr},
  {Bingham}, {Broos}, {Carter}, {Hancock}, {Huckle}, {Hunsberger}, {Kawakami},
  {Killough}, {Koch}, {McLelland}, {Smith}, {Smith}, {Soto}, {Boyd},
  {Breeveld}, {Holland}, {Ivanushkina}, {Pryzby}, {Still}, \&
  {Stock}}]{Roming2005}
{Roming}, P. W.~A., {Kennedy}, T.~E., {Mason}, K.~O., {et~al.} 2005, \ssr, 120,
  95, \dodoi{10.1007/s11214-005-5095-4}

\bibitem[{{Saz Parkinson} {et~al.}(2016){Saz Parkinson}, {Xu}, {Yu},
  {Salvetti}, {Marelli}, \& {Falcone}}]{SazParkinson2016}
{Saz Parkinson}, P.~M., {Xu}, H., {Yu}, P.~L.~H., {et~al.} 2016, \apj, 820, 8,
  \dodoi{10.3847/0004-637X/820/1/8}

\bibitem[{Wilms {et~al.}(2000)Wilms, Allen, \& Mccray}]{Wilms2000}
Wilms, J., Allen, A., \& Mccray, R. 2000, Astrophysical Journal, 542, 914

\bibitem[{{Wu} {et~al.}(2018){Wu}, {Clark}, {Pletsch}, {Guillemot}, {Johnson},
  {Torne}, {Champion}, {Deneva}, {Ray}, {Salvetti}, {Kramer}, {Aulbert},
  {Beer}, {Bhattacharyya}, {Bock}, {Camilo}, {Cognard}, {Cu{\'e}llar},
  {Eggenstein}, {Fehrmann}, {Ferrara}, {Kerr}, {Machenschalk}, {Ransom},
  {Sanpa-Arsa}, \& {Wood}}]{Wu2018}
{Wu}, J., {Clark}, C.~J., {Pletsch}, H.~J., {et~al.} 2018, \apj, 854, 99,
  \dodoi{10.3847/1538-4357/aaa411}

\bibitem[{{Zyuzin} {et~al.}(2018){Zyuzin}, {Karpova}, \&
  {Shibanov}}]{Zyuzin2018}
{Zyuzin}, D.~A., {Karpova}, A.~V., \& {Shibanov}, Y.~A. 2018, \mnras, 476,
  2177, \dodoi{10.1093/mnras/sty359}

\end{thebibliography}
\bibliographystyle{aasjournal}
\clearpage

\startlongtable
\begin{deluxetable*}{cccccc}
\label{tab:fitresults}
\tablecaption{X-ray photon indices and fluxes ($0.3-10.0 \: \rm{keV}$) for the 184 X-ray excesses matched with 3FGL unassociated sources used in this paper, including the five spectra unable to be fit by Xspec (marked with dashes) and the seven excluded spectra (\textbf{boldface}).  Certain spectra were unable to have asymmetric errors for $\Gamma_X$; these results use the default symmetric error and are marked with an asterisk (*). Available at CDS via anonymous ftp to cdsarc.u-strasbg.fr (130.79.128.5) or via \href{http://cdsarc.u-strasbg.fr/viz-bin/qcat?J/AJ}{http://cdsarc.u-strasbg.fr/viz-bin/qcat?J/AJ}}
\tablewidth{\columnwidth}
\tablehead{
\colhead{\textit{Swift} X-ray excess} & \colhead{3FGL gamma-ray source} &  \colhead{$\Gamma_{X}$} & \colhead{$\log_{10} F_X$} & \colhead{Cstat} & \colhead{DOF} \\
SwF3 & 3FGL & &  $\rm{erg/s/cm^2} $ & & } 
\startdata
J000805.3+145018 & J0008.3+1456 & $1.60^{0.15}_{0.14}$ & $-11.17^{+0.06}_{-0.05}$ & 82.17 & 89 \\
J003159.9+093615 & J0031.6+0938 & $2.11^{0.51}_{0.52}$ & $-12.41^{+0.17}_{-0.13}$ & 7.73 & 15 \\
J004859.5+422348 & J0049.0+4224 & $2.24^{0.38}_{0.39}$ & $-12.19^{+0.09}_{-0.08}$ & 33.91 & 34 \\
J012152.5$-$391544 & J0121.8$-$3917 & $1.81^{0.08}_{0.08}$ & $-11.45^{+0.03}_{-0.03}$ & 174.26 & 198 \\
J013255.1+593213 & J0133.3+5930 & $2.16^{0.24}_{0.23}$ & $-11.85^{+0.06}_{-0.06}$ & 51.59 & 57 \\
J015624.4$-$242003 & J0156.5$-$2423 & $2.30^{0.14}_{0.14}$ & $-11.90^{+0.04}_{-0.04}$ & 79.58 & 93 \\
J015852.4+010126 & J0158.6+0102 & $1.82^{0.82}_{0.73}$ & $-12.54^{+0.16}_{-0.17}$ & 4.09 & 6 \\
J020020.9$-$410933 & J0200.3$-$4108 & $2.66^{0.24}_{0.24}$ & $-12.19^{+0.06}_{-0.06}$ & 34.45 & 44 \\
J021210.6+532139 & J0212.1+5320 & $1.06^{0.14}_{0.14}$ & $-11.82^{+0.05}_{-0.05}$ & 73.84 & 110 \\
J023854.1+255405 & J0239.0+2555 & $2.17^{0.15}_{0.14}$ & $-11.83^{+0.04}_{-0.04}$ & 79.87 & 93 \\
J024454.9+475117 & J0244.4+4745 & $2.33^{0.50}_{0.50}$ & $-12.68^{+0.10}_{-0.11}$ & 14.22 & 14 \\
J025111.4$-$183115 & J0251.1$-$1829 & $2.01^{0.16}_{0.15}$ & $-12.20^{+0.05}_{-0.05}$ & 68.54 & 82 \\
J025857.4+055243 & J0258.9+0552 & $1.97^{0.30}_{0.29}$ & $-12.22^{+0.08}_{-0.08}$ & 23.23 & 31 \\
J034050.0$-$242259 & J0340.4$-$2423 & $2.27^{0.30}_{0.30}$ & $-12.50^{+0.09}_{-0.09}$ & 18.24 & 27 \\
J034158.1+314851 & J0342.3+3148c & $1.22^{0.16}_{0.16}$ & $-11.59^{+0.05}_{-0.05}$ & 90 & 97 \\
J034819.8+603506 & J0348.4+6039 & $1.42^{0.36}_{0.36}$ & $-11.09^{+0.09}_{-0.09}$ & 28.53 & 25 \\
J035051.2$-$281632 & J0351.0$-$2816 & $1.94^{0.06}_{0.06}$ & $-11.49^{+0.02}_{-0.02}$ & 235.76 & 239 \\
J035939.3+764627 & J0359.7+7649 & $3.10^{0.80}_{0.79}$ & $-12.13^{+0.21}_{-0.18}$ & 3.20 & 7 \\
J041433.2$-$084213 & J0414.9$-$0840 & $2.12^{0.38}_{0.39}$ & $-12.66^{+0.13}_{-0.10}$ & 21.33 & 25 \\
J042011.0$-$601504 & J0420.4$-$6013 & $2.15^{0.10}_{0.10}$ & $-11.69^{+0.03}_{-0.03}$ & 117.35 & 146 \\
J042749.8$-$670434* & J0427.9$-$6704 & $-0.15^{0.19}_{0.19}$ & $-11.78^{+0.06}_{-0.06}$ & 93.03 & 101 \\
J042958.7$-$305931 & J0430.1$-$3103 & $2.67^{0.24}_{0.24}$ & $-12.19^{+0.06}_{-0.06}$ & 41.66 & 42 \\
J043836.9$-$732919 & J0437.7$-$7330 & $2.08^{0.29}_{0.29}$ & $-12.39^{+0.08}_{-0.09}$ & 36.58 & 26 \\
J044722.5$-$253937 & J0447.1$-$2540 & $4.02^{0.93}_{0.75}$ & $-12.80^{+0.15}_{-0.16}$ & 4.52 & 8 \\
J045149.6+572140 & J0451.7+5722 & $2.82^{0.57}_{0.50}$ & $-12.42^{+0.18}_{-0.15}$ & 17.11 & 14 \\
J050650.1+032359 & J0506.9+0321 & $2.61^{0.14}_{0.13}$ & $-12.24^{+0.03}_{-0.03}$ & 96.36 & 117 \\
J051641.4+101243 & J0516.6+1012 & $2.32^{0.35}_{0.34}$ & $-12.40^{+0.08}_{-0.08}$ & 25.04 & 28 \\
J052939.5+382321 & J0529.2+3822 & $1.75^{0.46}_{0.45}$ & $-12.19^{+0.09}_{-0.09}$ & 27.12 & 24 \\
J053357.3$-$375754 & J0533.8$-$3754 & $2.50^{0.61}_{0.58}$ & $-12.39^{+0.13}_{-0.15}$ & 3.40 & 8 \\
J053559.3$-$061624 & J0535.7$-$0617c & $2.58^{0.32}_{0.32}$ & $-11.95^{+0.07}_{-0.07}$ & 28.41 & 34 \\
J055940.6+304232 & J0559.8+3042 & $1.92^{0.81}_{0.83}$ & $-12.36^{+0.22}_{-0.12}$ & 6.97 & 14 \\
J070421.7$-$482645 & J0704.3$-$4828 & $2.49^{0.15}_{0.14}$ & $-12.01^{+0.04}_{-0.04}$ & 78.40 & 99 \\
J071046.2$-$102942 & J0711.1$-$1037 & $1.04^{0.34}_{0.34}$ & $-11.73^{+0.10}_{-0.09}$ & 33.45 & 39 \\
J072547.5$-$054830 & J0725.7$-$0550 & $2.35^{0.16}_{0.16}$ & $-11.63^{+0.04}_{-0.04}$ & 91.01 & 105 \\
J074626.10$-$022551 & J0746.4$-$0225 & $2.41^{0.17}_{0.17}$ & $-11.87^{+0.04}_{-0.05}$ & 76.14 & 79 \\
J074724.8$-$492633 & J0747.5$-$4927 & $2.46^{0.22}_{0.21}$ & $-11.97^{+0.05}_{-0.05}$ & 55.18 & 56 \\
\textbf{J074903.8$-$221015} & J0748.8$-$2208 & $6.29^{1.86}_{1.76}$ & $-10.59^{+0.80}_{-0.77}$ & 4 & 4 \\
J080215.8$-$094214 & J0802.3$-$0941 & $2.44^{0.24}_{0.23}$ & $-12.20^{+0.06}_{-0.06}$ & 32.19 & 43 \\
J081338.1$-$035717 & J0813.5$-$0356 & $1.80^{0.12}_{0.12}$ & $-11.49^{+0.04}_{-0.04}$ & 142.84 & 125 \\
J082623.6$-$505742 & J0826.3$-$5056 & $1.35^{0.32}_{0.31}$ & $-12.27^{+0.09}_{-0.09}$ & 26.19 & 33 \\
J082628.2$-$640415 & J0826.3$-$6400 & $2.09^{0.05}_{0.05}$ & $-10.75^{+0.01}_{-0.01}$ & 312.05 & 327 \\
J083843.4$-$282701 & J0838.8$-$2829 & $1.75^{0.04}_{0.04}$ & $-11.25^{+0.01}_{-0.01}$ & 374.75 & 464 \\
J084831.8$-$694108 & J0847.2$-$6936 & $1.90^{0.21}_{0.21}$ & $-12.18^{+0.06}_{-0.06}$ & 67.42 & 66 \\
J085505.8$-$481517 & J0855.4$-$4818  & $-- \pm --$ & $--^{--}_{--}$ & $-$ & $-$ \\
\textbf{J090530.4$-$491840} & J0905.6$-$4917 & $7.84^{0.51}_{0.50}$ & $-8.99^{+0.23}_{-0.23}$ & 40.38 & 38 \\
J091926.1$-$220043 & J0919.5$-$2200 & $2.05^{0.20}_{0.20}$ & $-12.49^{+0.06}_{-0.06}$ & 48.26 & 52 \\
J092818.4$-$525659 & J0928.3$-$5255 & $2.98^{0.63}_{0.57}$ & $-11.47^{+0.33}_{-0.24}$ & 25.91 & 19 \\
J093444.6+090355 & J0935.2+0903 & $2.40^{0.79}_{0.81}$ & $-12.34^{+0.22}_{-0.17}$ & 7.13 & 6 \\
J093754.6$-$143349 & J0937.9$-$1435 & $2.42^{0.36}_{0.35}$ & $-12.53^{+0.08}_{-0.09}$ & 16.12 & 24 \\
J095249.5+071329 & J0952.8+0711 & $1.98^{0.24}_{0.23}$ & $-12.16^{+0.08}_{-0.08}$ & 38.43 & 37 \\
J101545.9$-$602938 & J1016.5$-$6034 & $3.08^{0.23}_{0.21}$ & $-11.54^{+0.09}_{-0.08}$ & 146.34 & 104 \\
J102432.6$-$454428 & J1024.4$-$4545 & $2.35^{0.12}_{0.11}$ & $-11.51^{+0.03}_{-0.03}$ & 112.07 & 145 \\
J103332.4$-$503526 & J1033.4$-$5035 & $2.28^{0.18}_{0.17}$ & $-11.75^{+0.04}_{-0.04}$ & 80.52 & 88 \\
J103755.1$-$242546 & J1038.0$-$2425 & $1.24^{0.26}_{0.27}$ & $-12.24^{+0.11}_{-0.11}$ & 34.05 & 31 \\
J103831.1$-$581346 & J1039.1$-$5809 & $4.28^{0.66}_{0.59}$ & $-11.38^{+0.36}_{-0.32}$ & 42.52 & 25 \\
J104939.4+154839 & J1049.7+1548 & $2.62^{0.17}_{0.17}$ & $-12.27^{+0.04}_{-0.04}$ & 68.58 & 73 \\
J105224.5+081409 & J1052.0+0816 & $1.90^{0.15}_{0.14}$ & $-11.62^{+0.05}_{-0.05}$ & 96.27 & 90 \\
J110025.5$-$205333 & J1100.2$-$2044 & $1.01^{0.66}_{0.79}$ & $-12.02^{+0.41}_{-0.30}$ & 14.86 & 11 \\
J110224.1$-$773339 & J1104.3$-$7736c & $2.47^{0.09}_{0.10}$ & $-11.77^{+0.02}_{-0.02}$ & 192.57 & 145 \\
J111601.8$-$484222 & J1116.7$-$4854 & $2.18^{0.41}_{0.41}$ & $-12.58^{+0.08}_{-0.08}$ & 29.45 & 32 \\
J111715.2$-$533815 & J1117.2$-$5338 & $1.97^{0.31}_{0.32}$ & $-12.08^{+0.08}_{-0.08}$ & 28.25 & 35 \\
J111956.10$-$264322 & J1119.8$-$2647 & $1.98^{0.32}_{0.32}$ & $-12.35^{+0.10}_{-0.09}$ & 37.97 & 27 \\
J111958.9$-$220456 & J1119.9$-$2204 & $1.89^{0.18}_{0.17}$ & $-13.10^{+0.06}_{-0.06}$ & 78.43 & 82 \\
J112042.4+071313 & J1120.6+0713 & $1.76^{0.26}_{0.26}$ & $-12.54^{+0.09}_{-0.09}$ & 36.75 & 29 \\
J112504.2$-$580539 & J1125.1$-$5803 & $2.47^{0.26}_{0.25}$ & $-11.66^{+0.07}_{-0.07}$ & 40.45 & 52 \\
J112624.8$-$500806 & J1126.8$-$5001 & $1.67^{0.06}_{0.07}$ & $-11.84^{+0.02}_{-0.02}$ & 248.78 & 282 \\
J113032.7$-$780107 & J1130.7$-$7800 & $1.80^{0.05}_{0.05}$ & $-10.80^{+0.02}_{-0.02}$ & 312.78 & 326 \\
J113209.3$-$473853 & J1132.0$-$4736 & $1.68^{0.07}_{0.07}$ & $-11.17^{+0.02}_{-0.02}$ & 268.25 & 287 \\
J114600.8$-$063851 & J1146.1$-$0640 & $2.12^{0.22}_{0.22}$ & $-12.01^{+0.07}_{-0.06}$ & 49.95 & 51 \\
J114911.10+280719 & J1149.1+2815 & $4.05^{0.77}_{0.62}$ & $-12.47^{+0.16}_{-0.14}$ & 13.54 & 10 \\
J115514.5$-$111125 & J1155.3$-$1112 & $1.78^{0.38}_{0.38}$ & $-12.28^{+0.15}_{-0.13}$ & 27.06 & 26 \\
J120055.1$-$143039 & J1200.9$-$1432 & $1.76^{0.26}_{0.26}$ & $-12.09^{+0.09}_{-0.08}$ & 39.89 & 37 \\
J121553.0$-$060940* & J1216.6$-$0557 & $1.50^{2.12}_{2.12}$ & $-12.37^{+16.37}_{-\infty}$ & 0 & 0 \\
J122014.4$-$245948 & J1220.0$-$2502 & $1.93^{0.11}_{0.11}$ & $-11.54^{+0.03}_{-0.03}$ & 114.51 & 151 \\
J122019.8$-$371414 & J1220.1$-$3715 & $2.05^{0.17}_{0.17}$ & $-11.78^{+0.05}_{-0.05}$ & 78.36 & 72 \\
J122127.4$-$062845 & J1221.5$-$0632 & $1.89^{0.17}_{0.16}$ & $-12.54^{+0.06}_{-0.06}$ & 58.62 & 73 \\
J122257.0+121438 & J1223.2+1215 & $1.89^{0.31}_{0.30}$ & $-13.09^{+0.10}_{-0.10}$ & 31.83 & 31 \\
J122536.7$-$344723 & J1225.4$-$3448 & $2.14^{0.10}_{0.10}$ & $-11.52^{+0.03}_{-0.03}$ & 116.66 & 154 \\
J123140.3+482148 & J1231.6+4825 & $1.97^{0.62}_{0.64}$ & $-12.38^{+0.25}_{-0.16}$ & 16.77 & 13 \\
J123204.2+165527 & J1232.3+1701  & $-- \pm --$ & $--^{--}_{--}$ & $-$ & $-$ \\
J123235.9$-$372055 & J1232.5$-$3720 & $2.40^{0.35}_{0.33}$ & $-12.40^{+0.09}_{-0.09}$ & 11.07 & 20 \\
J123447.7$-$043253 & J1234.7$-$0437 & $2.13^{0.39}_{0.38}$ & $-12.58^{+0.12}_{-0.12}$ & 19.29 & 16 \\
J123726.6$-$705140 & J1236.6$-$7050 & $1.67^{0.43}_{0.43}$ & $-12.20^{+0.13}_{-0.11}$ & 21.67 & 20 \\
J124021.3$-$714858 & J1240.3$-$7149 & $1.87^{0.05}_{0.05}$ & $-10.81^{+0.01}_{-0.01}$ & 310.77 & 370 \\
J124919.5$-$280833 & J1249.1$-$2808 & $2.08^{0.07}_{0.07}$ & $-11.46^{+0.02}_{-0.02}$ & 189.35 & 236 \\
J124919.7$-$054540 & J1249.5$-$0546 & $2.91^{0.31}_{0.30}$ & $-12.52^{+0.07}_{-0.08}$ & 20.84 & 31 \\
J125058.4$-$494444 & J1251.0$-$4943 & $2.32^{0.45}_{0.42}$ & $-12.52^{+0.09}_{-0.10}$ & 22.31 & 18 \\
J125821.5+212351 & J1258.4+2123 & $2.33^{0.21}_{0.21}$ & $-12.22^{+0.06}_{-0.06}$ & 43.69 & 48 \\
J130059.5$-$814809 & J1259.3$-$8151 & $1.29^{0.46}_{0.46}$ & $-12.31^{+0.17}_{-0.15}$ & 19.12 & 15 \\
J130128.9+333711 & J1301.5+3333 & $1.59^{0.40}_{0.40}$ & $-12.68^{+0.16}_{-0.15}$ & 17.50 & 16 \\
J130832.0+034406 & J1309.0+0347 & $1.59^{0.28}_{0.27}$ & $-12.13^{+0.11}_{-0.11}$ & 27.09 & 30 \\
J131140.3$-$623313* & J1311.8$-$6230 & $1.02^{1.51}_{1.51}$ & $-12.85^{+0.40}_{-0.34}$ & 1.84 & 2 \\
J131552.8$-$073304 & J1315.7$-$0732 & $2.37^{0.10}_{0.10}$ & $-11.64^{+0.03}_{-0.03}$ & 118.98 & 144 \\
J132928.6$-$053135 & J1329.1$-$0536 & $1.83^{0.13}_{0.13}$ & $-11.60^{+0.05}_{-0.05}$ & 92.98 & 98 \\
\textbf{J132939.6$-$610735} & J1329.8$-$6109 & $6.22^{0.54}_{0.59}$ & $-10.83^{+0.23}_{-0.26}$ & 26.73 & 22 \\
J140514.7$-$611822 & J1405.4$-$6119 & $-0.12^{0.37}_{0.38}$ & $-12.10^{+0.05}_{-0.05}$ & 129.53 & 128 \\
J141045.2+740504 & J1410.9+7406 & $2.59^{0.53}_{0.53}$ & $-12.79^{+0.12}_{-0.11}$ & 7.23 & 14 \\
J141133.3$-$072256 & J1411.4$-$0724 & $2.24^{0.26}_{0.25}$ & $-12.25^{+0.07}_{-0.07}$ & 30.32 & 35 \\
J142035.9$-$243021 & J1421.0$-$2431 & $4.79^{0.74}_{0.66}$ & $-12.06^{+0.16}_{-0.16}$ & 16.31 & 13 \\
J144544.5$-$593200 & J1445.7$-$5925 & $3.17^{1.07}_{0.96}$ & $-11.20^{+0.63}_{-0.44}$ & 5 & 5 \\
J151150.10+662450 & J1512.3+6622 & $1.74^{0.12}_{0.12}$ & $-11.70^{+0.05}_{-0.04}$ & 126.57 & 109 \\
J151256.6$-$564027* & J1512.8$-$5639 & $-0.35^{0.47}_{0.47}$ & $-11.70^{+0.12}_{-0.12}$ & 34.73 & 28 \\
J151319.0$-$372015 & J1513.3$-$3719 & $2.81^{0.28}_{0.28}$ & $-12.42^{+0.07}_{-0.07}$ & 40.28 & 34 \\
J151649.8+263635 & J1517.0+2637 & $2.20^{0.43}_{0.41}$ & $-12.59^{+0.10}_{-0.11}$ & 16.68 & 16 \\
J152603.0$-$083146 & J1525.8$-$0834  & $-- \pm --$ & $--^{--}_{--}$ & $-$ & $-$ \\
J152818.2$-$290256 & J1528.1$-$2904 & $1.72^{0.24}_{0.24}$ & $-12.10^{+0.08}_{-0.08}$ & 36.96 & 47 \\
J154150.1+141441 & J1541.6+1414 & $2.58^{0.44}_{0.37}$ & $-12.53^{+0.10}_{-0.10}$ & 18.66 & 18 \\
J154343.6$-$255607 & J1544.1$-$2555 & $3.10^{0.54}_{0.54}$ & $-12.34^{+0.12}_{-0.11}$ & 10.24 & 15 \\
J154459.2$-$664147 & J1545.0$-$6641 & $2.13^{0.06}_{0.06}$ & $-11.01^{+0.02}_{-0.02}$ & 269.72 & 300 \\
J154946.4$-$304502 & J1549.9$-$3044 & $2.24^{0.16}_{0.16}$ & $-11.80^{+0.04}_{-0.04}$ & 66.66 & 88 \\
J161543.2$-$444921 & J1615.6$-$4450 & $4.36^{0.42}_{0.44}$ & $-11.42^{+0.15}_{-0.16}$ & 44.98 & 23 \\
\textbf{J162432.2$-$465756} & J1624.1$-$4700 & $7.42^{0.42}_{0.42}$ & $-9.17^{+0.17}_{-0.18}$ & 32.20 & 40 \\
J162437.8$-$423144 & J1624.8$-$4233 & $3.08^{0.65}_{0.53}$ & $-12.09^{+0.22}_{-0.18}$ & 27.59 & 14 \\
J162607.8$-$242736 & J1626.2$-$2428c & $0.87^{0.56}_{0.56}$ & $-12.34^{+0.17}_{-0.16}$ & 8.14 & 12 \\
J162743.0+322102 & J1627.8+3217 & $2.38^{0.20}_{0.19}$ & $-12.47^{+0.05}_{-0.06}$ & 51.73 & 53 \\
J165338.2$-$015837 & J1653.6$-$0158 & $1.29^{0.28}_{0.28}$ & $-12.73^{+0.10}_{-0.10}$ & 38.62 & 37 \\
J170409.6+123423 & J1704.1+1234 & $2.02^{0.10}_{0.09}$ & $-11.50^{+0.03}_{-0.03}$ & 137.89 & 163 \\
J170433.9$-$052840 & J1704.4$-$0528 & $1.91^{0.09}_{0.09}$ & $-11.43^{+0.03}_{-0.03}$ & 144.39 & 197 \\
J170521.6$-$413436 & 1710.6$-$4128c & $-- \pm --$ & $--^{--}_{--}$ & $-$ & $-$ \\
\textbf{J171106.10$-$432415} & J1710.6$-$4317 & $6.32^{0.42}_{0.43}$ & $-10.26^{+0.16}_{-0.18}$ & 40.58 & 39 \\
J172142.1$-$392204 & J1721.8$-$3919 & $1.08^{0.24}_{0.25}$ & $-11.78^{+0.06}_{-0.06}$ & 63.43 & 65 \\
J172858.2+604359 & J1729.0+6049 & $4.79^{1.55}_{1.42}$ & $-11.72^{+0.57}_{-0.45}$ & 16.32 & 7 \\
J173250.5+591233 & J1732.7+5914 & $2.14^{0.30}_{0.29}$ & $-12.36^{+0.08}_{-0.08}$ & 20.59 & 28 \\
J173508.1$-$292955 & J1734.7$-$2930 & $1.94^{0.41}_{0.40}$ & $-11.91^{+0.09}_{-0.09}$ & 23.55 & 23 \\
J174511.10$-$225455 & J1744.7$-$2252 & $0.88^{0.14}_{0.15}$ & $-11.74^{+0.05}_{-0.05}$ & 111.59 & 129 \\
J175316.4$-$444822 & J1753.6$-$4447 & $2.03^{0.40}_{0.41}$ & $-12.58^{+0.10}_{-0.10}$ & 13.72 & 19 \\
J175359.7$-$292909 & J1754.0$-$2930 & $0.83^{0.22}_{0.22}$ & $-11.80^{+0.07}_{-0.07}$ & 71.97 & 74 \\
J180351.7+252607 & J1804.1+2532 & $1.20^{0.35}_{0.34}$ & $-12.39^{+0.13}_{-0.12}$ & 23.03 & 22 \\
J180425.0$-$085002 & J1804.5$-$0850 & $1.48^{0.33}_{0.32}$ & $-12.08^{+0.09}_{-0.09}$ & 23.34 & 31 \\
J181307.6$-$684713 & J1813.6$-$6845 & $2.04^{0.46}_{0.46}$ & $-12.95^{+0.13}_{-0.12}$ & 25.63 & 20 \\
J181720.4$-$303257 & J1817.3$-$3033 & $1.86^{0.13}_{0.13}$ & $-11.80^{+0.04}_{-0.04}$ & 108.11 & 129 \\
J182914.0+272901 & J1829.2+2731 & $1.25^{0.19}_{0.19}$ & $-12.10^{+0.07}_{-0.07}$ & 50.73 & 65 \\
J182915.5+323432 & J1829.2+3229 & $2.46^{0.83}_{0.73}$ & $-12.02^{+0.22}_{-0.18}$ & 4.32 & 6 \\
J183659.5$-$240027 & J1837.3$-$2403 & $1.17^{0.93}_{1.04}$ & $-13.28^{+0.31}_{-0.27}$ & 5.42 & 7 \\
J184433.1$-$034627* & J1844.3$-$0344 & $0.62^{0.64}_{0.64}$ & $-12.64^{+0.12}_{-0.11}$ & 31.22 & 29 \\
J184833.10+323249 & J1848.6+3232 & $0.47^{0.62}_{0.65}$ & $-12.45^{+0.25}_{-0.24}$ & 10.01 & 8 \\
J185520.0+075138 & J1855.6+0753 & $0.84^{0.40}_{0.40}$ & $-12.12^{+0.11}_{-0.10}$ & 43.66 & 26 \\
J185606.6$-$122147 & J1856.1$-$1217 & $2.06^{0.26}_{0.25}$ & $-11.78^{+0.05}_{-0.06}$ & 45.45 & 52 \\
J190444.5$-$070743 & J1904.7$-$0708 & $2.21^{0.35}_{0.34}$ & $-12.23^{+0.08}_{-0.09}$ & 25.28 & 26 \\
\textbf{J192113.10+194004} & J1921.6+1934 & $6.19^{0.51}_{0.49}$ & $-10.13^{+0.22}_{-0.22}$ & 41.87 & 30 \\
J192242.1$-$745354 & J1923.2$-$7452 & $2.25^{0.09}_{0.09}$ & $-11.44^{+0.03}_{-0.03}$ & 167.01 & 174 \\
J193420.1+600138 & J1934.2+6002 & $2.51^{0.27}_{0.27}$ & $-12.11^{+0.06}_{-0.06}$ & 33.01 & 45 \\
J194633.6$-$540235 & J1946.4$-$5403 & $1.69^{0.24}_{0.24}$ & $-12.68^{+0.09}_{-0.09}$ & 39.86 & 39 \\
J195149.7+690719 & J1951.3+6909 & $2.65^{0.47}_{0.45}$ & $-12.33^{+0.09}_{-0.09}$ & 16.30 & 19 \\
J195800.3+243803 & J1958.1+2436 & $2.22^{0.31}_{0.31}$ & $-11.43^{+0.09}_{-0.07}$ & 33.81 & 48 \\
J200635.7+015222 & J2006.6+0150 & $1.82^{0.32}_{0.32}$ & $-12.40^{+0.11}_{-0.10}$ & 28.78 & 22 \\
J201020.3$-$212434 & J2010.0$-$2120 & $1.50^{0.50}_{0.52}$ & $-12.14^{+0.21}_{-0.18}$ & 14.71 & 11 \\
J201525.3$-$143204 & J2015.3$-$1431 & $3.15^{0.35}_{0.34}$ & $-12.28^{+0.08}_{-0.08}$ & 19.68 & 28 \\
J203027.9$-$143919 & J2030.5$-$1439 & $1.51^{0.32}_{0.30}$ & $-12.19^{+0.12}_{-0.12}$ & 21.77 & 31 \\
J203450.9$-$420038 & J2034.6$-$4202 & $2.59^{0.12}_{0.12}$ & $-11.84^{+0.03}_{-0.03}$ & 108.50 & 121 \\
\textbf{J203556.9+490038} & J2035.8+4902 & $5.82^{0.61}_{0.63}$ & $-10.70^{+0.24}_{-0.26}$ & 21.96 & 21 \\
J203935.8+123001 & J2039.7+1237  & $-- \pm --$ & $--^{--}_{--}$ & $-$ & $-$ \\
J204351.5+103407 & J2044.0+1035 & $2.51^{0.33}_{0.33}$ & $-12.34^{+0.08}_{-0.08}$ & 21.71 & 28 \\
J204806.3$-$312011 & J2047.9$-$3119 & $2.21^{0.28}_{0.28}$ & $-12.23^{+0.07}_{-0.07}$ & 28.19 & 38 \\
J205350.8+292312 & J2053.9+2922 & $2.24^{0.09}_{0.09}$ & $-10.93^{+0.02}_{-0.02}$ & 191.26 & 194 \\
J205357.9+690517 & J2054.3+6907 & $0.54^{0.31}_{0.32}$ & $-12.66^{+0.11}_{-0.12}$ & 38.95 & 32 \\
J205950.4+202905 & J2059.9+2029 & $2.22^{0.35}_{0.34}$ & $-12.28^{+0.08}_{-0.09}$ & 26.63 & 23 \\
J210940.0+043958 & J2110.0+0442 & $1.99^{0.25}_{0.24}$ & $-11.94^{+0.07}_{-0.07}$ & 36.05 & 39 \\
J211522.2+121801 & J2115.2+1215 & $3.18^{0.47}_{0.45}$ & $-12.38^{+0.12}_{-0.11}$ & 8.35 & 18 \\
J212051.6$-$125300 & J2120.4$-$1256 & $0.66^{0.39}_{0.41}$ & $-12.57^{+0.16}_{-0.16}$ & 7.79 & 17 \\
J212601.5+583148 & J2125.8+5832 & $1.81^{0.36}_{0.35}$ & $-12.08^{+0.08}_{-0.08}$ & 35.73 & 26 \\
J212729.3$-$600102 & J2127.5$-$6001 & $1.24^{0.15}_{0.13}$ & $-11.48^{+0.05}_{-0.06}$ & 131.95 & 94 \\
J214247.5+195811 & J2142.7+1957 & $2.38^{0.22}_{0.22}$ & $-11.83^{+0.05}_{-0.05}$ & 64.31 & 60 \\
J214429.5$-$563849 & J2144.6$-$5640 & $2.04^{0.25}_{0.24}$ & $-12.27^{+0.08}_{-0.08}$ & 35.85 & 38 \\
J215046.5$-$174956 & J2150.5$-$1754 & $3.50^{1.06}_{0.75}$ & $-12.38^{+0.19}_{-0.16}$ & 13.51 & 9 \\
J215122.10+415634 & J2151.6+4154 & $2.58^{0.24}_{0.23}$ & $-11.61^{+0.06}_{-0.06}$ & 42.19 & 63 \\
J220941.4$-$045108 & J2209.8$-$0450 & $2.46^{0.49}_{0.51}$ & $-12.53^{+0.10}_{-0.11}$ & 12.92 & 13 \\
J222911.2+225456 & J2229.1+2255 & $2.14^{0.07}_{0.07}$ & $-11.24^{+0.02}_{-0.02}$ & 191.20 & 215 \\
J224437.0+250344 & J2244.6+2503 & $2.87^{0.36}_{0.34}$ & $-12.42^{+0.08}_{-0.09}$ & 39.92 & 24 \\
J224710.1$-$000512* & J2247.2$-$0004 & $0.30^{0.60}_{0.60}$ & $-12.57^{+0.33}_{-0.30}$ & 13.32 & 8 \\
J225003.5$-$594520 & J2249.3$-$5943 & $1.71^{0.41}_{0.42}$ & $-12.40^{+0.17}_{-0.14}$ & 13.36 & 18 \\
J225032.7+174918 & J2250.3+1747 & $1.26^{0.42}_{0.43}$ & $-12.64^{+0.18}_{-0.17}$ & 14.05 & 13 \\
J225045.7+330514 & J2250.6+3308 & $1.63^{0.65}_{0.68}$ & $-12.42^{+0.28}_{-0.19}$ & 10.01 & 12 \\
J230012.4+405223 & J2300.0+4053 & $1.99^{0.09}_{0.09}$ & $-11.20^{+0.03}_{-0.03}$ & 135.26 & 181 \\
J230351.7+555617 & J2303.7+5555 & $1.84^{0.17}_{0.17}$ & $-11.53^{+0.04}_{-0.04}$ & 77.52 & 99 \\
J230848.5+542612 & J2309.0+5428 & $2.54^{0.27}_{0.26}$ & $-12.19^{+0.07}_{-0.07}$ & 47.02 & 46 \\
J232127.1+511117 & J2321.3+5113 & $2.55^{0.40}_{0.40}$ & $-12.22^{+0.09}_{-0.09}$ & 27.99 & 24 \\
J232137.1$-$161926 & J2321.6$-$1619 & $2.45^{0.12}_{0.12}$ & $-11.61^{+0.03}_{-0.03}$ & 101.96 & 119 \\
J232653.3$-$412713 & J2327.2$-$4130 & $1.97^{0.41}_{0.39}$ & $-12.48^{+0.12}_{-0.12}$ & 19.61 & 16 \\
J232938.7+610111 & J2329.8+6102 & $2.62^{0.18}_{0.18}$ & $-11.28^{+0.06}_{-0.06}$ & 94.22 & 120 \\
J233626.4$-$842649 & J2337.2$-$8425 & $2^{0.24}_{0.24}$ & $-12.12^{+0.07}_{-0.07}$ & 35.91 & 47 \\
J235115.9$-$760017 & J2351.9$-$7601 & $1.99^{0.21}_{0.20}$ & $-12.03^{+0.06}_{-0.06}$ & 56.64 & 51 \\
J235824.10+382857 & J2358.5+3827 & $2.15^{0.15}_{0.15}$ & $-11.68^{+0.04}_{-0.04}$ & 86 & 96 \\
J235836.8$-$180717 & J2358.6$-$1809 & $2.57^{0.10}_{0.10}$ & $-11.55^{+0.03}_{-0.03}$ & 125.82 & 135 \\
\enddata
\end{deluxetable*}

\newcolumntype{H}{>{\setbox0=\hbox\bgroup}c<{\egroup}@{}}

\begin{longrotatetable}
\begin{deluxetable*}{ccccccc}
\label{tab:MLresults}
\tablecaption{RF classification results for the unassociated sources and paired excesses. We used the grid-search optimized RF parameters to determine the blazar probabilities. Available at CDS via anonymous ftp to cdsarc.u-strasbg.fr (130.79.128.5) or via \href{http://cdsarc.u-strasbg.fr/viz-bin/qcat?J/AJ}{http://cdsarc.u-strasbg.fr/viz-bin/qcat?J/AJ}}
\tablehead{
\colhead{\textit{Swift} X-ray excess} & \colhead{3FGL gamma-ray source} & \colhead{X-ray excess RA} & \colhead{X-ray excess Decl} & \colhead{Blazar Prob.} &  \colhead{Categorization} \\
SwF3 & 3FGL & (J2000) & (J2000) & $P_{bzr}$ &
} 
\startdata
J000805.3+145018 & J0008.3+1456 & 00 08 05.27 & +14 50 18.5 & 0.976 & blazar \\
J003159.9+093615 & J0031.6+0938 & 00 31 59.89 & +09 36 15.9 & 0.979 & blazar \\
J004859.5+422348 & J0049.0+4224 & 00 48 59.45 & +42 23 48.8 & 0.990 & blazar \\
J012152.5$-$391544 & J0121.8$-$3917 & 01 21 52.53 & $-$39 15 44.7 & 0.943 & blazar \\
J013255.1+593213 & J0133.3+5930 & 01 32 55.11 & +59 32 13.3 & 0.999 & blazar \\
J015624.4$-$242003 & J0156.5$-$2423 & 01 56 24.44 & $-$24 20 03.4 & 1.000 & blazar \\
J015852.4+010126 & J0158.6+0102 & 01 58 52.44 & +01 01 26.9 & 0.988 & blazar \\
J020020.9$-$410933 & J0200.3$-$4108 & 02 00 20.90 & $-$41 09 33.9 & 1.000 & blazar \\
J021210.6+532139 & J0212.1+5320 & 02 12 10.55 & +53 21 39.6 & 0.081 & pulsar \\
J023854.1+255405 & J0239.0+2555 & 02 38 54.11 & +25 54 05.6 & 0.991 & blazar \\
J024454.9+475117 & J0244.4+4745 & 02 44 54.93 & +47 51 17.5 & 0.817 &  \\
J025111.4$-$183115 & J0251.1$-$1829 & 02 51 11.39 & $-$18 31 15.1 & 0.966 & blazar \\
J025857.4+055243 & J0258.9+0552 & 02 58 57.44 & +05 52 43.7 & 0.996 & blazar \\
J034050.0$-$242259 & J0340.4$-$2423 & 03 40 50.02 & $-$24 22 59.2 & 0.995 & blazar \\
J034158.1+314851 & J0342.3+3148c & 03 41 58.10 & +31 48 51.7 & 0.841 &  \\
J034819.8+603506 & J0348.4+6039 & 03 48 19.77 & +60 35 06.6 & 0.986 & blazar \\
J035051.2$-$281632 & J0351.0$-$2816 & 03 50 51.24 & $-$28 16 32.6 & 0.998 & blazar \\
J035939.3+764627 & J0359.7+7649 & 03 59 39.29 & +76 46 27.5 & 0.999 & blazar \\
J041433.2$-$084213 & J0414.9$-$0840 & 04 14 33.20 & $-$08 42 13.2 & 0.997 & blazar \\
J042011.0$-$601504 & J0420.4$-$6013 & 04 20 11.02 & $-$60 15 04.8 & 0.998 & blazar \\
J042749.8$-$670434 & J0427.9$-$6704 & 04 27 49.76 & $-$67 04 34.8 & 0.772 &  \\
J042958.7$-$305931 & J0430.1$-$3103 & 04 29 58.74 & $-$30 59 31.7 & 0.989 & blazar \\
J043836.9$-$732919 & J0437.7$-$7330 & 04 38 36.85 & $-$73 29 19.9 & 0.994 & blazar \\
J044722.5$-$253937 & J0447.1$-$2540 & 04 47 22.52 & $-$25 39 37.4 & 0.998 & blazar \\
J045149.6+572140 & J0451.7+5722 & 04 51 49.56 & +57 21 40.6 & 0.993 & blazar \\
J050650.1+032359 & J0506.9+0321 & 05 06 50.08 & +03 23 59.6 & 0.999 & blazar \\
J051641.4+101243 & J0516.6+1012 & 05 16 41.44 & +10 12 43.4 & 0.984 & blazar \\
J052939.5+382321 & J0529.2+3822 & 05 29 39.54 & +38 23 21.4 & 0.777 &  \\
J053357.3$-$375754 & J0533.8$-$3754 & 05 33 57.33 & $-$37 57 54.6 & 0.920 & blazar \\
J053559.3$-$061624 & J0535.7$-$0617c & 05 35 59.28 & $-$06 16 24.1 & 0.999 & blazar \\
J055940.6+304232 & J0559.8+3042 & 05 59 40.58 & +30 42 32.7 & 0.985 & blazar \\
J070421.7$-$482645 & J0704.3$-$4828 & 07 04 21.74 & $-$48 26 45.3 & 0.990 & blazar \\
J071046.2$-$102942 & J0711.1$-$1037 & 07 10 46.18 & $-$10 29 42.2 & 0.437 &  \\
J072547.5$-$054830 & J0725.7$-$0550 & 07 25 47.51 & $-$05 48 30.3 & 0.997 & blazar \\
J074626.10$-$022551 & J0746.4$-$0225 & 07 46 26.98 & $-$02 25 51.8 & 0.965 & blazar \\
J074724.8$-$492633 & J0747.5$-$4927 & 07 47 24.78 & $-$49 26 33.9 & 0.999 & blazar \\
J080215.8$-$094214 & J0802.3$-$0941 & 08 02 15.78 & $-$09 42 14.0 & 0.998 & blazar \\
J081338.1$-$035717 & J0813.5$-$0356 & 08 13 38.10 & $-$03 57 17.1 & 0.968 & blazar \\
J082623.6$-$505742 & J0826.3$-$5056 & 08 26 23.55 & $-$50 57 42.5 & 0.806 &  \\
J082628.2$-$640415 & J0826.3$-$6400 & 08 26 28.16 & $-$64 04 15.5 & 0.941 & blazar \\
J083843.4$-$282701 & J0838.8$-$2829 & 08 38 43.37 & $-$28 27 01.6 & 0.719 &  \\
J084831.8$-$694108 & J0847.2$-$6936 & 08 48 31.82 & $-$69 41 08.9 & 0.994 & blazar \\
J085505.8$-$481517 & J0855.4$-$4818 & 08 55 05.78 & $-$48 15 17.9 & 0.130 &  \\
J091926.1$-$220043 & J0919.5$-$2200 & 09 19 26.09 & $-$22 00 43.6 & 0.995 & blazar \\
J092818.4$-$525659 & J0928.3$-$5255 & 09 28 18.40 & $-$52 56 59.7 & 0.980 & blazar \\
J093444.6+090355 & J0935.2+0903 & 09 34 44.60 & +09 03 55.8 & 0.807 &  \\
J093754.6$-$143349 & J0937.9$-$1435 & 09 37 54.55 & $-$14 33 49.1 & 0.999 & blazar \\
J095249.5+071329 & J0952.8+0711 & 09 52 49.52 & +07 13 29.6 & 0.996 & blazar \\
J101545.9$-$602938 & J1016.5$-$6034 & 10 15 45.85 & $-$60 29 38.1 & 0.213 &  \\
J102432.6$-$454428 & J1024.4$-$4545 & 10 24 32.56 & $-$45 44 28.5 & 1.000 & blazar \\
J103332.4$-$503526 & J1033.4$-$5035 & 10 33 32.38 & $-$50 35 26.7 & 1.000 & blazar \\
J103755.1$-$242546 & J1038.0$-$2425 & 10 37 55.09 & $-$24 25 46.0 & 0.981 & blazar \\
J103831.1$-$581346 & J1039.1$-$5809 & 10 38 31.12 & $-$58 13 46.6 & 0.497 &  \\
J104939.4+154839 & J1049.7+1548 & 10 49 39.44 & +15 48 39.0 & 0.973 & blazar \\
J105224.5+081409 & J1052.0+0816 & 10 52 24.46 & +08 14 09.5 & 0.994 & blazar \\
J110025.5$-$205333 & J1100.2$-$2044 & 11 00 25.46 & $-$20 53 33.1 & 0.906 & blazar \\
J110224.1$-$773339 & J1104.3$-$7736c & 11 02 24.13 & $-$77 33 39.9 & 0.916 & blazar \\
J111601.8$-$484222 & J1116.7$-$4854 & 11 16 01.82 & $-$48 42 22.6 & 0.966 & blazar \\
J111715.2$-$533815 & J1117.2$-$5338 & 11 17 15.15 & $-$53 38 15.3 & 0.997 & blazar \\
J111956.10$-$264322 & J1119.8$-$2647 & 11 19 56.96 & $-$26 43 22.2 & 0.995 & blazar \\
J111958.9$-$220456 & J1119.9$-$2204 & 11 19 58.93 & $-$22 04 56.6 & 0.032 & pulsar \\
J112042.4+071313 & J1120.6+0713 & 11 20 42.35 & +07 13 13.1 & 0.337 &  \\
J112504.2$-$580539 & J1125.1$-$5803 & 11 25 04.21 & $-$58 05 39.6 & 1.000 & blazar \\
J112624.8$-$500806 & J1126.8$-$5001 & 11 26 24.82 & $-$50 08 06.8 & 0.991 & blazar \\
J113032.7$-$780107 & J1130.7$-$7800 & 11 30 32.65 & $-$78 01 07.5 & 0.902 & blazar \\
J113209.3$-$473853 & J1132.0$-$4736 & 11 32 09.31 & $-$47 38 53.8 & 0.981 & blazar \\
J114600.8$-$063851 & J1146.1$-$0640 & 11 46 00.77 & $-$06 38 51.0 & 0.998 & blazar \\
J114911.10+280719 & J1149.1+2815 & 11 49 11.98 & +28 07 19.9 & 0.985 & blazar \\
J115514.5$-$111125 & J1155.3$-$1112 & 11 55 14.54 & $-$11 11 25.3 & 0.896 &  \\
J120055.1$-$143039 & J1200.9$-$1432 & 12 00 55.07 & $-$14 30 39.0 & 0.992 & blazar \\
J121553.0$-$060940 & J1216.6$-$0557 & 12 15 53.04 & $-$06 09 40.6 & 0.988 & blazar \\
J122014.4$-$245948 & J1220.0$-$2502 & 12 20 14.43 & $-$24 59 48.0 & 0.994 & blazar \\
J122019.8$-$371414 & J1220.1$-$3715 & 12 20 19.82 & $-$37 14 14.1 & 0.999 & blazar \\
J122127.4$-$062845 & J1221.5$-$0632 & 12 21 27.38 & $-$06 28 45.9 & 0.989 & blazar \\
J122257.0+121438 & J1223.2+1215 & 12 22 57.04 & +12 14 38.5 & 0.986 & blazar \\
J122536.7$-$344723 & J1225.4$-$3448 & 12 25 36.66 & $-$34 47 23.8 & 0.999 & blazar \\
J123140.3+482148 & J1231.6+4825 & 12 31 40.28 & +48 21 48.9 & 0.993 & blazar \\
J123204.2+165527 & J1232.3+1701 & 12 32 04.22 & +16 55 27.7 & 0.776 &  \\
J123235.9$-$372055 & J1232.5$-$3720 & 12 32 35.87 & $-$37 20 55.8 & 0.992 & blazar \\
J123447.7$-$043253 & J1234.7$-$0437 & 12 34 47.69 & $-$04 32 53.9 & 0.976 & blazar \\
J123726.6$-$705140 & J1236.6$-$7050 & 12 37 26.61 & $-$70 51 40.1 & 0.996 & blazar \\
J124021.3$-$714858 & J1240.3$-$7149 & 12 40 21.34 & $-$71 48 58.3 & 0.973 & blazar \\
J124919.5$-$280833 & J1249.1$-$2808 & 12 49 19.46 & $-$28 08 33.5 & 0.999 & blazar \\
J124919.7$-$054540 & J1249.5$-$0546 & 12 49 19.69 & $-$05 45 40.2 & 1.000 & blazar \\
J125058.4$-$494444 & J1251.0$-$4943 & 12 50 58.43 & $-$49 44 44.3 & 0.998 & blazar \\
J125821.5+212351 & J1258.4+2123 & 12 58 21.49 & +21 23 51.7 & 0.865 &  \\
J130059.5$-$814809 & J1259.3$-$8151 & 13 00 59.45 & $-$81 48 09.6 & 0.992 & blazar \\
J130128.9+333711 & J1301.5+3333 & 13 01 28.85 & +33 37 11.0 & 0.793 &  \\
J130832.0+034406 & J1309.0+0347 & 13 08 32.00 & +03 44 06.9 & 0.779 &  \\
J131140.3$-$623313 & J1311.8$-$6230 & 13 11 40.25 & $-$62 33 13.9 & 0.577 &  \\
J131552.8$-$073304 & J1315.7$-$0732 & 13 15 52.84 & $-$07 33 04.0 & 0.998 & blazar \\
J132928.6$-$053135 & J1329.1$-$0536 & 13 29 28.56 & $-$05 31 35.4 & 0.995 & blazar \\
J140514.7$-$611822 & J1405.4$-$6119 & 14 05 14.67 & $-$61 18 22.7 & 0.113 &  \\
J141045.2+740504 & J1410.9+7406 & 14 10 45.24 & +74 05 04.8 & 0.849 &  \\
J141133.3$-$072256 & J1411.4$-$0724 & 14 11 33.30 & $-$07 22 56.3 & 0.996 & blazar \\
J142035.9$-$243021 & J1421.0$-$2431 & 14 20 35.85 & $-$24 30 21.9 & 0.901 & blazar \\
J144544.5$-$593200 & J1445.7$-$5925 & 14 45 44.49 & $-$59 32 00.2 & 0.996 & blazar \\
J151150.10+662450 & J1512.3+6622 & 15 11 50.95 & +66 24 50.1 & 0.988 & blazar \\
J151256.6$-$564027 & J1512.8$-$5639 & 15 12 56.57 & $-$56 40 27.3 & 0.865 &  \\
J151319.0$-$372015 & J1513.3$-$3719 & 15 13 19.03 & $-$37 20 15.0 & 0.977 & blazar \\
J151649.8+263635 & J1517.0+2637 & 15 16 49.82 & +26 36 35.2 & 0.995 & blazar \\
J152603.0$-$083146 & J1525.8$-$0834 & 15 26 03.03 & $-$08 31 46.2 & 0.785 &  \\
J152818.2$-$290256 & J1528.1$-$2904 & 15 28 18.18 & $-$29 02 56.8 & 0.981 & blazar \\
J154150.1+141441 & J1541.6+1414 & 15 41 50.12 & +14 14 41.2 & 0.983 & blazar \\
J154343.6$-$255607 & J1544.1$-$2555 & 15 43 43.60 & $-$25 56 07.6 & 0.804 &  \\
J154459.2$-$664147 & J1545.0$-$6641 & 15 44 59.17 & $-$66 41 47.8 & 0.991 & blazar \\
J154946.4$-$304502 & J1549.9$-$3044 & 15 49 46.41 & $-$30 45 02.3 & 0.883 &  \\
J161543.2$-$444921 & J1615.6$-$4450 & 16 15 43.18 & $-$44 49 21.1 & 0.966 & blazar \\
J162437.8$-$423144 & J1624.8$-$4233 & 16 24 37.81 & $-$42 31 44.3 & 0.986 & blazar \\
J162607.8$-$242736 & J1626.2$-$2428c & 16 26 07.79 & $-$24 27 36.2 & 0.330 &  \\
J162743.0+322102 & J1627.8+3217 & 16 27 43.00 & +32 21 02.4 & 0.776 &  \\
J165338.2$-$015837 & J1653.6$-$0158 & 16 53 38.15 & $-$01 58 37.0 & 0.045 & pulsar \\
J170409.6+123423 & J1704.1+1234 & 17 04 09.60 & +12 34 23.0 & 0.977 & blazar \\
J170433.9$-$052840 & J1704.4$-$0528 & 17 04 33.91 & $-$05 28 40.7 & 0.986 & blazar \\
J170521.6$-$413436 & J1705.5$-$4128c & 17 05 21.56 & $-$41 34 36.9 & 0.749 &  \\
J172142.1$-$392204 & J1721.8$-$3919 & 17 21 42.13 & $-$39 22 04.7 & 0.929 & blazar \\
J172858.2+604359 & J1729.0+6049 & 17 28 58.17 & +60 43 59.9 & 0.995 & blazar \\
J173250.5+591233 & J1732.7+5914 & 17 32 50.51 & +59 12 33.8 & 0.999 & blazar \\
J173508.1$-$292955 & J1734.7$-$2930 & 17 35 08.07 & $-$29 29 55.4 & 0.496 &  \\
J174511.10$-$225455 & J1744.7$-$2252 & 17 45 11.98 & $-$22 54 55.7 & 0.177 &  \\
J175316.4$-$444822 & J1753.6$-$4447 & 17 53 16.43 & $-$44 48 22.0 & 0.668 &  \\
J175359.7$-$292909 & J1754.0$-$2930 & 17 53 59.69 & $-$29 29 09.0 & 0.165 &  \\
J180351.7+252607 & J1804.1+2532 & 18 03 51.72 & +25 26 07.0 & 0.780 &  \\
J180425.0$-$085002 & J1804.5$-$0850 & 18 04 25.02 & $-$08 50 02.5 & 0.966 & blazar \\
J181307.6$-$684713 & J1813.6$-$6845 & 18 13 07.62 & $-$68 47 13.2 & 0.909 & blazar \\
J181720.4$-$303257 & J1817.3$-$3033 & 18 17 20.35 & $-$30 32 57.8 & 0.985 & blazar \\
J182914.0+272901 & J1829.2+2731 & 18 29 14.00 & +27 29 01.8 & 0.713 &  \\
J182915.5+323432 & J1829.2+3229 & 18 29 15.50 & +32 34 32.4 & 0.851 &  \\
J183659.5$-$240027 & J1837.3$-$2403 & 18 36 59.46 & $-$24 00 27.9 & 0.455 &  \\
J184433.1$-$034627 & J1844.3$-$0344 & 18 44 33.14 & $-$03 46 27.4 & 0.020 & pulsar \\
J184833.10+323249 & J1848.6+3232 & 18 48 33.96 & +32 32 49.6 & 0.503 &  \\
J185520.0+075138 & J1855.6+0753 & 18 55 20.02 & +07 51 38.6 & 0.747 &  \\
J185606.6$-$122147 & J1856.1$-$1217 & 18 56 06.60 & $-$12 21 47.7 & 0.894 &  \\
J190444.5$-$070743 & J1904.7$-$0708 & 19 04 44.53 & $-$07 07 43.1 & 0.968 & blazar \\
J192242.1$-$745354 & J1923.2$-$7452 & 19 22 42.10 & $-$74 53 54.9 & 0.980 & blazar \\
J193420.1+600138 & J1934.2+6002 & 19 34 20.08 & +60 01 38.2 & 0.999 & blazar \\
J194633.6$-$540235 & J1946.4$-$5403 & 19 46 33.57 & $-$54 02 35.1 & 0.042 & pulsar \\
J195149.7+690719 & J1951.3+6909 & 19 51 49.66 & +69 07 19.2 & 0.971 & blazar \\
J195800.3+243803 & J1958.1+2436 & 19 58 00.28 & +24 38 03.8 & 0.997 & blazar \\
J200635.7+015222 & J2006.6+0150 & 20 06 35.73 & +01 52 22.4 & 0.861 &  \\
J201020.3$-$212434 & J2010.0$-$2120 & 20 10 20.25 & $-$21 24 34.2 & 0.737 &  \\
J201525.3$-$143204 & J2015.3$-$1431 & 20 15 25.25 & $-$14 32 04.7 & 0.999 & blazar \\
J203027.9$-$143919 & J2030.5$-$1439 & 20 30 27.92 & $-$14 39 19.0 & 0.997 & blazar \\
J203450.9$-$420038 & J2034.6$-$4202 & 20 34 50.88 & $-$42 00 38.1 & 0.988 & blazar \\
J203935.8+123001 & J2039.7+1237 & 20 39 35.76 & +12 30 01.8 & 0.719 &  \\
J204351.5+103407 & J2044.0+1035 & 20 43 51.54 & +10 34 07.7 & 0.921 & blazar \\
J204806.3$-$312011 & J2047.9$-$3119 & 20 48 06.25 & $-$31 20 11.5 & 0.901 & blazar \\
J205350.8+292312 & J2053.9+2922 & 20 53 50.77 & +29 23 12.4 & 0.914 & blazar \\
J205357.9+690517 & J2054.3+6907 & 20 53 57.94 & +69 05 17.7 & 0.787 &  \\
J205950.4+202905 & J2059.9+2029 & 20 59 50.44 & +20 29 05.0 & 0.984 & blazar \\
J210940.0+043958 & J2110.0+0442 & 21 09 40.04 & +04 39 58.3 & 1.000 & blazar \\
J211522.2+121801 & J2115.2+1215 & 21 15 22.17 & +12 18 01.7 & 1.000 & blazar \\
J212051.6$-$125300 & J2120.4$-$1256 & 21 20 51.55 & $-$12 53 00.4 & 0.851 &  \\
J212601.5+583148 & J2125.8+5832 & 21 26 01.49 & +58 31 48.3 & 0.630 &  \\
J212729.3$-$600102 & J2127.5$-$6001 & 21 27 29.30 & $-$60 01 02.1 & 0.970 & blazar \\
J214247.5+195811 & J2142.7+1957 & 21 42 47.47 & +19 58 11.9 & 1.000 & blazar \\
J214429.5$-$563849 & J2144.6$-$5640 & 21 44 29.48 & $-$56 38 49.7 & 0.918 & blazar \\
J215046.5$-$174956 & J2150.5$-$1754 & 21 50 46.46 & $-$17 49 56.0 & 0.913 & blazar \\
J215122.10+415634 & J2151.6+4154 & 21 51 22.99 & +41 56 34.5 & 0.999 & blazar \\
J220941.4$-$045108 & J2209.8$-$0450 & 22 09 41.35 & $-$04 51 08.2 & 0.933 & blazar \\
J222911.2+225456 & J2229.1+2255 & 22 29 11.17 & +22 54 56.1 & 0.998 & blazar \\
J224437.0+250344 & J2244.6+2503 & 22 44 37.00 & +25 03 44.3 & 0.930 & blazar \\
J224710.1$-$000512 & J2247.2$-$0004 & 22 47 10.11 & $-$00 05 12.1 & 0.488 &  \\
J225003.5$-$594520 & J2249.3$-$5943 & 22 50 03.45 & $-$59 45 20.4 & 0.993 & blazar \\
J225032.7+174918 & J2250.3+1747 & 22 50 32.71 & +17 49 18.4 & 0.980 & blazar \\
J225045.7+330514 & J2250.6+3308 & 22 50 45.65 & +33 05 14.8 & 0.647 &  \\
J230012.4+405223 & J2300.0+4053 & 23 00 12.36 & +40 52 23.3 & 0.992 & blazar \\
J230351.7+555617 & J2303.7+5555 & 23 03 51.70 & +55 56 17.9 & 0.968 & blazar \\
J230848.5+542612 & J2309.0+5428 & 23 08 48.49 & +54 26 12.0 & 1.000 & blazar \\
J232127.1+511117 & J2321.3+5113 & 23 21 27.12 & +51 11 17.8 & 1.000 & blazar \\
J232137.1$-$161926 & J2321.6$-$1619 & 23 21 37.05 & $-$16 19 26.2 & 0.999 & blazar \\
J232653.3$-$412713 & J2327.2$-$4130 & 23 26 53.32 & $-$41 27 13.4 & 0.993 & blazar \\
J232938.7+610111 & J2329.8+6102 & 23 29 38.70 & +61 01 11.6 & 1.000 & blazar \\
J233626.4$-$842649 & J2337.2$-$8425 & 23 36 26.35 & $-$84 26 49.5 & 0.989 & blazar \\
J235115.9$-$760017 & J2351.9$-$7601 & 23 51 15.89 & $-$76 00 17.6 & 0.994 & blazar \\
J235824.10+382857 & J2358.5+3827 & 23 58 24.95 & +38 28 57.2 & 1.000 & blazar \\
J235836.8$-$180717 & J2358.6$-$1809 & 23 58 36.75 & $-$18 07 17.9 & 1.000 & blazar \\
\enddata
\end{deluxetable*}
\end{longrotatetable}

\end{document}